\documentclass[PRD,reprint,onecolumn,showpacs,
 nofootinbib, superscriptaddress]{revtex4}

\usepackage{color}
\usepackage{amsmath,amssymb}
\usepackage[colorlinks,allcolors=blue]{hyperref}
\usepackage{graphicx}
\usepackage{dcolumn}
\usepackage{bm}
\usepackage{url}

\newcommand{\beq}{\begin{equation}}
\newcommand{\eeq}{\end{equation}}
\newcommand{\beqa}{\begin{eqnarray}}
\newcommand{\eeqa}{\end{eqnarray}}

\begin{document}

\title{GeV excess in the Milky Way: \\The Role of Diffuse Galactic gamma ray Emission template}
\affiliation{Key Laboratory of Dark Matter and Space Astronomy, Purple Mountain Observatory, Chinese Academy of Sciences, Nanjing 210008, China}
\affiliation{University of Chinese Academy of Sciences, Yuquan Road 19, Beijing, 100049, China.}
\author{Bei Zhou}
\affiliation{Key Laboratory of Dark Matter and Space Astronomy, Purple Mountain Observatory, Chinese Academy of Sciences, Nanjing 210008, China}
\affiliation{University of Chinese Academy of Sciences, Yuquan Road 19, Beijing, 100049, China.}
\author{Yun-Feng Liang}
\affiliation{Key Laboratory of Dark Matter and Space Astronomy, Purple Mountain Observatory, Chinese Academy of Sciences, Nanjing 210008, China}
\affiliation{University of Chinese Academy of Sciences, Yuquan Road 19, Beijing, 100049, China.}
\author{Xiaoyuan Huang$^\ast$}
\affiliation{Key Laboratory of Dark Matter and Space Astronomy, Purple Mountain Observatory, Chinese Academy of Sciences, Nanjing 210008, China}
\author{Xiang Li$^\ast$}
\affiliation{Key Laboratory of Dark Matter and Space Astronomy, Purple Mountain Observatory, Chinese Academy of Sciences, Nanjing 210008, China}
\affiliation{University of Chinese Academy of Sciences, Yuquan Road 19, Beijing, 100049, China.}
\author{Yi-Zhong Fan$^\ast$}
\affiliation{Key Laboratory of Dark Matter and Space Astronomy, Purple Mountain Observatory, Chinese Academy of Sciences, Nanjing 210008, China}
\author{Lei Feng}
\affiliation{Key Laboratory of Dark Matter and Space Astronomy, Purple Mountain Observatory, Chinese Academy of Sciences, Nanjing 210008, China}
\author{Jin Chang}
\affiliation{Key Laboratory of Dark Matter and Space Astronomy, Purple Mountain Observatory, Chinese Academy of Sciences, Nanjing 210008, China}
\date{\today}

\begin{abstract}
Several groups have analyzed the publicly-available Fermi-LAT data and reported a spatially extended $\gamma-$ray excess of around $1-3$ GeV  from the region surrounding the Galactic Center that might originate from annihilation of dark matter particles with a rest mass $m_\chi \sim 30-40$ GeV. In this work we examine the role of the diffuse Galactic gamma ray emission (DGE) templates played in suppressing the GeV excess. For such a purpose, we adopt in total 128 background templates that have been generated by Ackermann et al. \cite{FermiLAT:2012aa} in the study of the {Fermi-LAT} observations of the diffuse gamma ray emission considering the effects of cosmic rays and the interstellar medium. The possible GeV excess, assumed to follow the spatial distribution of the prompt gamma-rays produced in the annihilation of dark matter particles taking a generalized NFW profile with an inner slope $\alpha=1.2$, has been analyzed in some regions of interest. The introduction of such an additional component centered at the Galactic center is found to have improved the goodness of fit to the data significantly in all background template models regardless of whether the excess spectrum is fixed or not. Our results thus suggest that the presence of a statistically significant GeV excess in the inner Galaxy is robust thought its spectrum depends on the DGE model adopted in the analysis. The possible physical origin of the GeV excess component is discussed and in the dark matter model the annihilation cross section of such particles is evaluated.
\end{abstract}

\pacs{98.70.Rz, 95.35.+d}

\maketitle

\section{Introduction}
In the standard $\Lambda$CDM cosmology model, the current universe consists of $\sim 4.9\%$ baryonic matter, $\sim 26.8\%$ cold dark matter and $\sim 68.3\%$ dark energy \cite{Ade:2013zuv}. Though abundant, the nature of dark matter particles is still poorly understood. Among various viable dark matter candidates, weakly interacting massive particles (WIMPs) have been the most extensively discussed and are suggested to be the leading ones \cite{Jungman:1995df, Bertone:2004pz, Hooper:2007qk, Feng:2010gw, Fan:2010yq}. WIMPs may be able to annihilate with each other (or alternatively decay) and then produce energetic particles, including gamma rays, charged particles, and neutrinos. Thanks to the specific radiation spectra of such components, the dark matter-originated gamma-rays or/and cosmic rays may be identifiable from the dense astrophysical background. The cosmic rays are deflected by the magnetic fields and lose energies before reaching us. As a result, the direction information is lost and the dark matter origin of some cosmic ray anomaly, for example the electrons/positrons excesses \cite{Adriani:2008zr, Aguilar:2013qda, Adriani:2011xv, Chang:2008aa, FermiLAT:2011ab}, is somewhat challenging to establish (sometimes the large uncertainty of the cosmic ray background is also an obstacle \cite{Li:2014csu}). The prompt photons from the annihilation (or decay) events instead trace the dark matter distribution. The morphology of the gamma ray signal is hence valuable for establishing the dark matter origin.
The Galactic Center, the dwarf galaxies and the galaxy clusters are the regions of interest for dark matter indirect detection in gamma rays. While for most of the dwarf galaxies and the galaxy clusters they can only be observed as point sources and the morphology information is missed. The Galactic Center, benefited from its proximity and high dark matter density, is expected to be the brightest prompt photon source of dark matter annihilation on the sky, and an spatial extension of the dark matter annihilation signal is expected. Since the launch of the Fermi Gamma-Ray Space Telescope \cite{Atwood:2007ra, Atwood:2009ez}, many groups have studied the possible dark matter induced signal in the Galactic Center. One tentative signal is a monochromatic gamma ray line with energy $\sim$130 GeV \cite{Bringmann:2012vr, Weniger:2012tx, Tempel:2012ey, Su:2012ft, Yang:2013kra, Ackermann:2013uma}. Another interesting signal is a statistically-very-important GeV excess  concentrating at the Galactic Center \cite{Goodenough:2009gk, Vitale:2009hr, Hooper:2010mq, Hooper:2010im, Abazajian:2012pn, Gordon:2013vta, Huang:2013pda} but extending to a Galactic latitude $|b| \sim 10^\circ-20^\circ$ \cite{Hooper:2013rwa}. Both the spectrum and the morphology of the GeV excess component in the  Galactic Center and
Inner Galaxy are found to be compatible with that predicted from the annihilations of WIMPs with a rest mass $\sim 30-40$ GeV via the channels mainly to quarks \cite{Daylan:2014rsa}. Together with the electron/positron data, the annihilation channels can be further constrained. For example, the dark matter annihilates to combination of channels, with cross sections proportional to the square of the charge of the final state particles or democratically to all kinetically allowed standard model fermions (note that the quark final states win by additional factor 3 from color), are found to be ruled out \cite{Huang2014}.

The progress on identifying a possible GeV excess centered at the Galactic center was remarkable in the past few years. And statistically the significance of the GeV excess is so high that it is unlikely to be a fluctuation. Nevertheless, the role of the  diffuse Galactic gamma ray emission (DGE) template in suppressing the GeV excess signal is to be carefully examined. For such a purpose, in this work, following \cite{FermiLAT:2012aa} we adopt in total 128 background templates that have been used in the gamma ray study of the Fermi-LAT observations of the DGE considering the effects of cosmic rays and the interstellar medium. These diffuse Galactic gamma ray emission background templates
(models), created by varying within observational limits the distribution
of cosmic-ray sources, the size of the cosmic-ray confinement volume, and the distribution of interstellar
gas, are constrained by local cosmic rays observations \cite{FermiLAT:2012aa}. With each template we evaluate the statistical significance of the possible GeV excess component in some regions of interest.  The excess component has been assumed to follow the spatial distribution of the gamma-rays produced in the annihilation of dark matter particles taking a generalized NFW profile\cite{Navarro:1995iw, Navarro:1996gj} with an inner slope $\alpha=1.2$.

This work is structured as the following. In section 2 we briefly introduce the background templates used in the data analysis and also the regions of interest. In section 3 we present the method and results of our data analysis. In section 4 we summarize our results with some discussion on the prospect of confirming or ruling out the dark matter annihilation origin of the GeV excess.

\section{The diffuse Galactic gamma ray emission templates and regions of interest used in this analysis}\label{Sec:Templates}
Cosmic rays propagating through the Milky Way interact with interstellar gas, magnetic fields as well as the soft photons, and then generate the observed DGE that dominates in the energy range of Fermi-LAT. In the search of signal from large-scale regions using
Fermi-LAT data, it is essential to know the DGE well to get a robust result unless the target signal has very special spectral features \cite{Bringmann:2012vr, Weniger:2012tx, Tempel:2012ey, Su:2012ft, Yang:2013kra,  Ackermann:2013uma}.
But in reality, the distribution of cosmic rays, interstellar gas, magnetic fields as well as radiation fields are still not precisely known.  Correspondingly, the predicted DGE suffers from some uncertainties, which in turn weakens the robustness of the observed signal. Such a fact motivates us to investigate the role of DGE templates in shaping the GeV excess signal reported in the literature.

DGE can be divided into three components based on its origin \cite{FermiLAT:2012aa}: (a) hadronic
emission from neutral pion decay produced by inelastic collision of cosmic ray protons with the interstellar gas, (b) inverse Compton
scattering of interstellar soft photons by cosmic ray electrons and positrons, and (c) bremsstrahlung produced by scattering of
cosmic ray electrons and positrons with protons/nuclei in interstellar gas. The neutral pion decay and bremsstrahlung emission are both generated
by cosmic ray particles interacting with interstellar gas, and their spatial distributions will both follow the morphology of the target gas. That is why in some approaches, for example to have the Fermi LAT source catalog \cite{Abdo:2010ru, Fermi-LAT:2011iqa},  the spatial templates of interstellar gas have been used to fit with Fermi-LAT data directly to get the spectral energy distribution and to model the effect of pion decay and bremsstrahlung emission, with the assumption that the cosmic rays flux is uniform within each template. However, the spectral information of these two kinds of radiation processes have not been taken into account and the constraints from local cosmic ray observations have been ignored.
The templates generated in such a way are optimized for point sources and small scale extended sources and are not ideal templates for analyzing  spatially extended sources and/or large-scale diffuse emission \cite{DIFFUSE2009, DIFFUSE2013}\footnote{\url{http://fermi.gsfc.nasa.gov/ssc/data/analysis/LAT\_caveats.html}}.

One way to get both the spatial and spectral information of DGE templates is to adopt the GALPROP code
\cite{Strong:1998pw} to calculate the propagation and distribution of cosmic rays in the Milky Way, and then the radiation of
cosmic rays interacting with interstellar gas and radiation fields \cite{FermiLAT:2012aa}. The cosmic ray sources are essentially unknown. In  \cite{FermiLAT:2012aa}, four kinds of cosmic ray source distribution models (SNR distribution \cite{Case:1998qg}, Lorimer pulsar distribution \cite{Lorimer:2006qs}, Yusifov pulsar distribution \cite{Yusifov:2004fr} and OB stars distribution \cite{Bronfman:2000tw}) were adopted, and eight combinations of different sizes for the cosmic ray confinement region were used, namely two radial boundaries (i.e., $R_{\rm h}$=20 and 30 kpc) and four vertical boundaries (i.e., $z_{\rm h}$=4, 6, 8, and 10 kpc), respectively. The cosmic ray diffusion
equations were solved and the cosmic ray injection and diffusion parameters
were inferred from fitting the local cosmic ray observation data.
There were some additional assumptions on the column density of the gas, two for spin temperature ($T_{\rm s}=150$ K and $10^{5}$ K, respectively)
which could affect the derived HI column densities and two for dust ($E(B-V)=2$ mag and 5 mag, respectively) as the tracer of gas. We refer the readers to Section 3 of \cite{FermiLAT:2012aa} for the details of these model input parameters. Finally, interactions of cosmic
rays with targets were calculated to predict the induced gamma ray distribution and an all sky fit to the Fermi-LAT data
was performed to determine the rest parameters. In total 128 DGE models were created, which will be used in our template-dependent GeV excess analysis. In this work we use the supplementary online material \footnote{\url{http://galprop.stanford.edu/PaperIISuppMaterial/}} to generate templates in mapCube format which can be easily convolved with
Point Spread Function (PSF) \footnote{\url{http://fermi.gsfc.nasa.gov/ssc/data/analysis/documentation/Cicerone/Cicerone_LAT_IRFs/IRF_PSF.html}} of Fermi-LAT using Fermi Science Tools \footnote{\url{http://fermi.gsfc.nasa.gov/ssc/data/analysis/software/}}. In Table 3 of \cite{FermiLAT:2012aa}, the model numbers were given to DGE models with different sets of specific parameters. In our analysis we take the same approach.

Galactic center hosts a lot of sources with energetic activities that could be accelerators of cosmic rays,
for example the supermassive black hole Sgr A* and supernova remnants. Young populations of cosmic rays
likely inhabit in the Milky Way center and give rise to the hadronic emission. These gamma ray emissions are unpredicted in
the templates generated in \cite{Carlson:2014cwa} because these possible accelerators
 have been excluded in the four cosmic ray source distribution models. In principle such an ``ignored" hadronic emission component could contribute to the GeV excess. Moreover, the interaction between high energy electrons and molecular clouds in the galactic center
 produce significant bremsstrahlung radiation. These emissions may partially contribute to the central region GeV excess too \cite{YusefZadeh:2012nh}.
In addition to these unpredicted radiations, millisecond pulsars, which are suggested to be abundant
in the Milky Way center and unresolvable because of limited PSF of Fermi-LAT, can generate extended gamma ray emission with a
peak around GeV \cite{Abazajian:2012pn, Gordon:2013vta, Mirabal:2013rba, Yuan:2014rca, Calore:2014oga}. Please also bear in mind that the flux and spectrum of central point source 2FGL J1745.6-2858 associated with Sgr A* have some degeneracies with the GeV excess \cite{Abazajian:2012pn}. Therefore the Galactic center is not a perfect region to establish the dark matter or other novel origin of the GeV excess signal because many astrophysical processes could generate similar diffusion emission. In view of these complications, we should optimize our region of interest (ROI) to reduce the uncertainties from background modeling and then reliably identify additional signal(s).

\begin{figure}
    \begin{center}
        \includegraphics[width=0.48\linewidth]{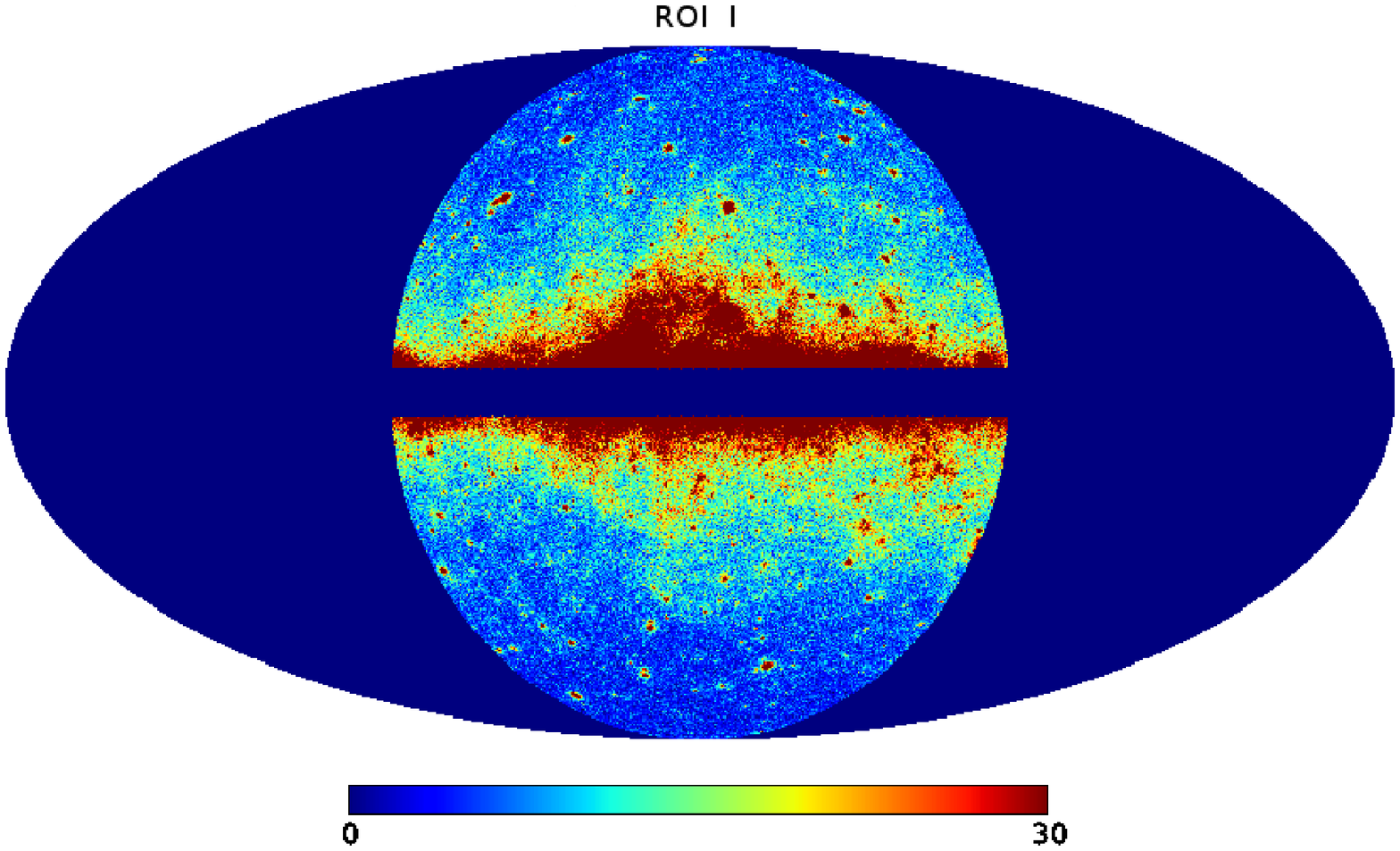}\\
        \includegraphics[width=0.48\linewidth]{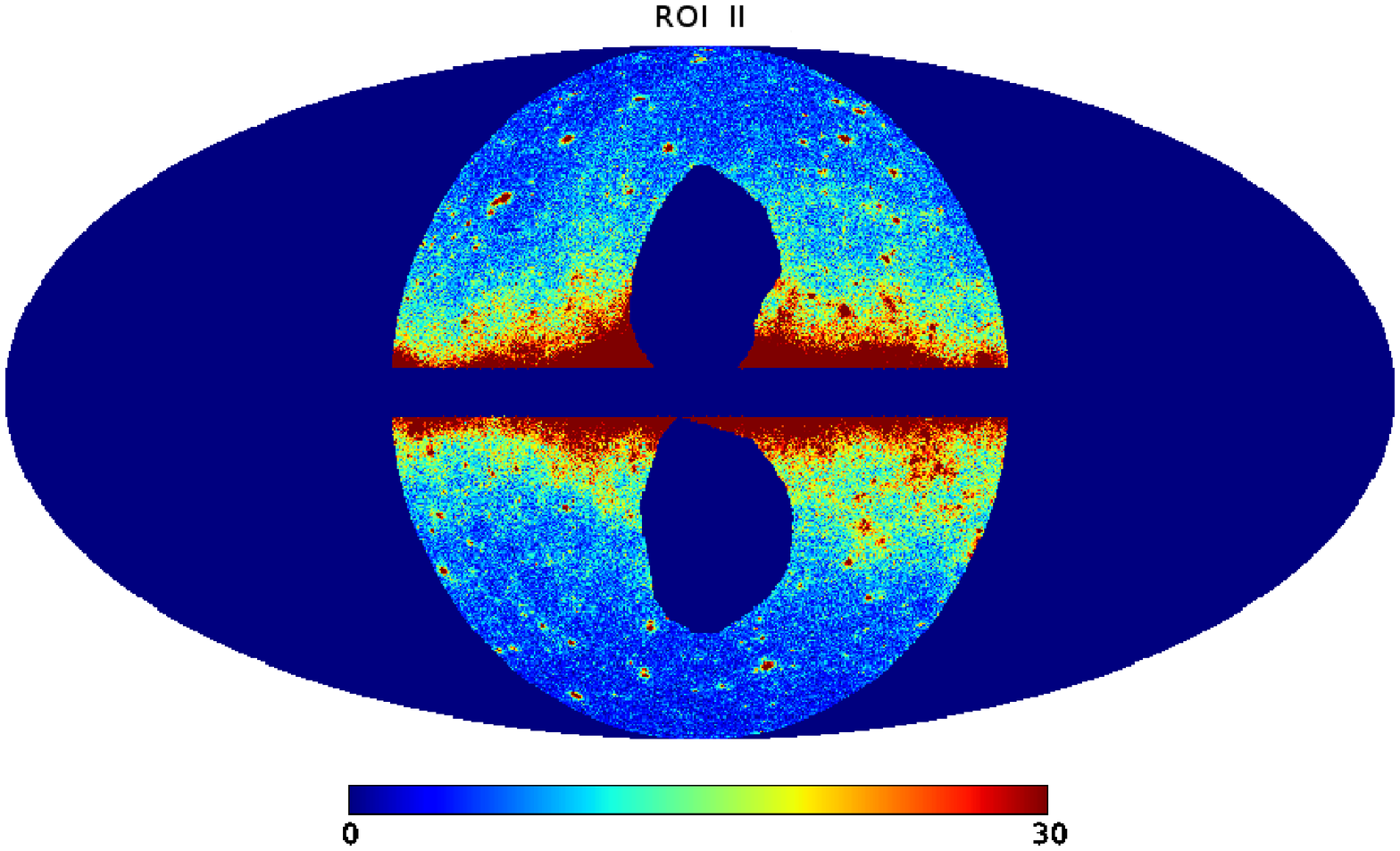}
        \includegraphics[width=0.48\linewidth]{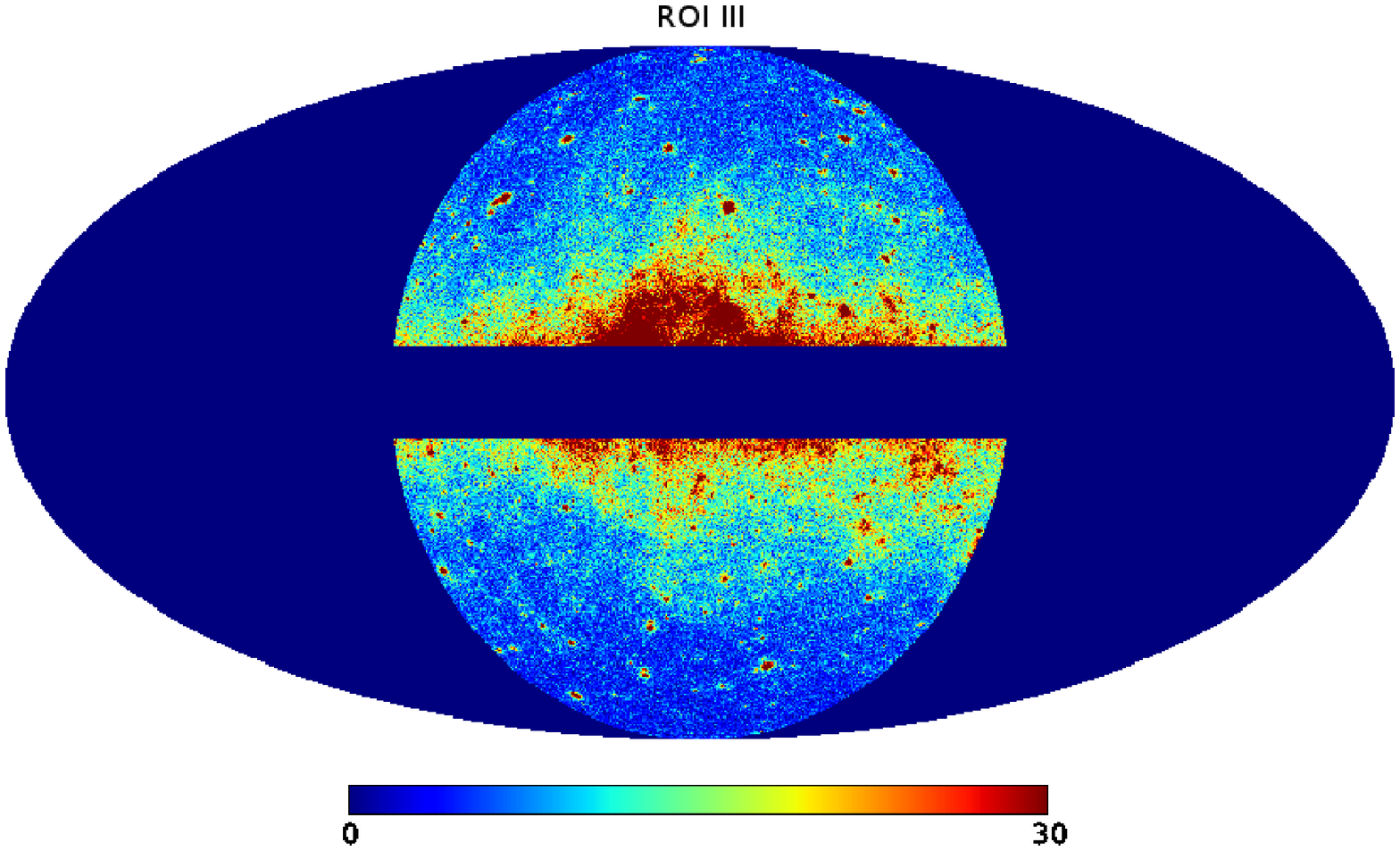}
    \end{center}
    \caption{The regions of interest are chosen to minimize the uncertainties from background modeling. What shown here are the count maps of gamma-rays in the energy range of $300~{\rm MeV}-300~{\rm GeV}$.  In the top panel (i.e., ROI I), we mask  $\left| b \right| <5^\circ$ to reduce the contamination from known under-predicted gamma ray emission above a few GeV in the Galactic plane \cite{FermiLAT:2012aa} and also to avoid analyzing the complex Galactic center. We also mask $\left| l \right| >80^\circ$ to  minimize the
effect of outer Galaxy region which dominates in an all sky likelihood ratio test but will not considerably contribute to the dark matter annihilation signal \cite{FermiLAT:2012aa}. In the bottom/left panel (i.e., ROI II), the Fermi Bubbles are further masked to minimize its possible degeneracy with dark matter signal in low latitude region. In the bottom/right panel, we mask the regions of  $\left| b \right| <10^\circ$ and $\left| l \right| >80^\circ$ (i.e., ROI III) to minimize the contamination of possible misleading GeV emission,  which would be less extended, from millisecond pulsars,pulsars, SNRs, bremsstrahlung or neutral pion decay.
}
\label{fig:masked}
\end{figure}

It was aware that in \cite{FermiLAT:2012aa} the created DGE models always under-predict the gamma ray emission above a few GeV in the Galactic plane, possibly because the contribution from unresolved point sources such as pulsars, SNRs and pulsar wind nebulae has not been taken into account.
Then in Sec.\ref{sec:IIIB} we mask the $|b|<5^\circ$ and $|b|<10^\circ$ regions respectively to minimize the possible ``misleading" components from Galactic center and Galactic plane.  The likelihood fit in \cite{FermiLAT:2012aa} found out that the outer Galaxy would dominate in an all sky likelihood ratio test. But if the GeV excess signal is from dark matter annihilation, we would anticipate that the outer Galaxy region will not contribute significantly to the signal. That is why we also mask the region of $|l| >80^\circ$ to minimize the effect of outer Galaxy region that will dilute the `potential' GeV excess signal. The regions of ($|b| <5^\circ$, $|l| >80^\circ$) and ($|b| <10^\circ$, $|l| >80^\circ$) are our ROI I and ROI III, respectively.  Hooper et al. \cite{Hooper:2013rwa} found that the Fermi Bubbles \cite{Su:2010qj} might have a uniform brightness intensity as long as a proper dark matter-like additional component has been taken into account. But the astrophysical origin of the Fermi Bubbles is unknown and the Fermi Bubbles may be nonuniform but with some hot spots \cite{Su:2010qj, Su:2012gu}. As found in the SED analysis (see Fig.~\ref{fig:p6v11}),  the low latitude region of the Fermi Bubbles has a degeneracy with the dark matter template in the low energy range. In order to test the possible connection between the GeV excess and the Fermi
Bubbles, we also select a region of interest as $|l|<80^\circ$ and $|b| >5^\circ$ excluding Fermi Bubbles (i.e., ROI II).
In Sec.\ref{sec:IIIB} we carry out the data analysis in all these three regions of interest (see Fig.~\ref{fig:masked}).
While in Sec.\ref{sec:IIIC} the data analysis is performed just in ROI I.

\section{Data analysis: method and results}
\subsection{Methodology}\label{sec:IIIA}

We use the public gamma ray data of Fermi-LAT from 300 MeV to 300 GeV between August 4, 2008 (MET 239557417) and April 7, 2014 (MET 418537497).
The ULTRACLEAN data set was selected to reduce contamination from the charged particles.
We also employ standard cuts for diffuse analysis including zenith angle $< 100^\circ$, {DATA$_{-}$QUAL} = 1, {LAT$_{-}$CONFIG} = 1 and instrumental rocking angle (i.e. angle of the spacecraft $Z-$axis from zenith) $<52^\circ$.

The data were divided into 30 logarithmic energy bins and we make the maps of counts into each energy bin for FRONT and BACK events respectively to HEALPIX grids with NSIDE=256.
The BACK events in the first 6 energy bins were ignored in our analysis because of their low quality especially for the PSF that is considerably worse than that of the FRONT events. Therefore in the following procedures, we always do analysis for FRONT and BACK maps separately.

Although the effect of point sources is subordinate for large sky region analysis (see Appendix \ref{app:src}), to make our analysis more robust, we take into account the point and extended sources in the Fermi-LAT 2-year catalog\footnote{\url{http://fermi.gsfc.nasa.gov/ssc/data/access/lat/2yr_catalog/}} (2FGL) \cite{Fermi-LAT:2011iqa}  and add them into our model in the subsequent data fitting (for the details see Appendix \ref{app:likelihhood}).
The parameters of these sources are fixed to save the computational time.
In view of that the 2-year catalog data are likely unable to accurately represent the average flux of the sources in 5.5 years especially for the brightest sources, we mask the brightest 200 sources throughout our analysis and the sizes of masked regions are determined by Fermi-LAT PSF.

The templates employed in our fits incorporate
(a) A group of galactic diffuse emission models, each contains three components accounting for bremsstrahlung, $\pi^0$ decay as well as the inverse Compton radiation of the Galaxy respectively. In total we have 128 groups of such models, as mentioned in Sec.\ref{Sec:Templates}.
(b) An uniform-brightness template of Fermi Bubbles defined by \cite{Su:2010qj};
(c) a dark-matter-annihilation-like spatial distribution template defined by the generalized NFW profile $\rho \propto (r/r_{\rm s})^{-\alpha}(1+r/r_{\rm s})^{-3+\alpha}$ \cite{Navarro:1995iw, Navarro:1996gj}, where $r_{\rm s}=20$ kpc is the scale radius and  $\alpha=$1.2 is the slope index \footnote{The slope index is fixed otherwise it is very time-consuming.};
(d) a collection of all point/extended sources in the LAT 2-year catalog;
(e) an isotropic map used to absorb residual cosmic-ray contamination and isotropic diffuse emission.
Furthermore, we convolve the templates with the Fermi-LAT PSF to match the data in each energy bin for FRONT and BACK events, respectively.
As found in \cite{Daylan:2014rsa}, the galactic center GeV excess can be well explained by the dark matter particles with the mass of $30-40$ GeV annihilating to $b\bar{b}$ pairs. So in our likelihood fit of the GeV excess in Sec. \ref{sec:IIIB}, the $\gamma-$ray spectrum of $\sim 35$ GeV dark matter particles annihilating to $b\bar{b}$ will be adopted. While in Sec. \ref{sec:IIIC} the spectrum of the `potential' excess is not given as a priority any longer.
We fit the maps of counts with linear combinations of the 5 sets of templates, maximizing the pixel-based Poisson likelihood, and each time we just change (a), i.e., the galactic diffuse emission model.

\begin{figure}
    \begin{center}
        \includegraphics[width=0.48\linewidth]{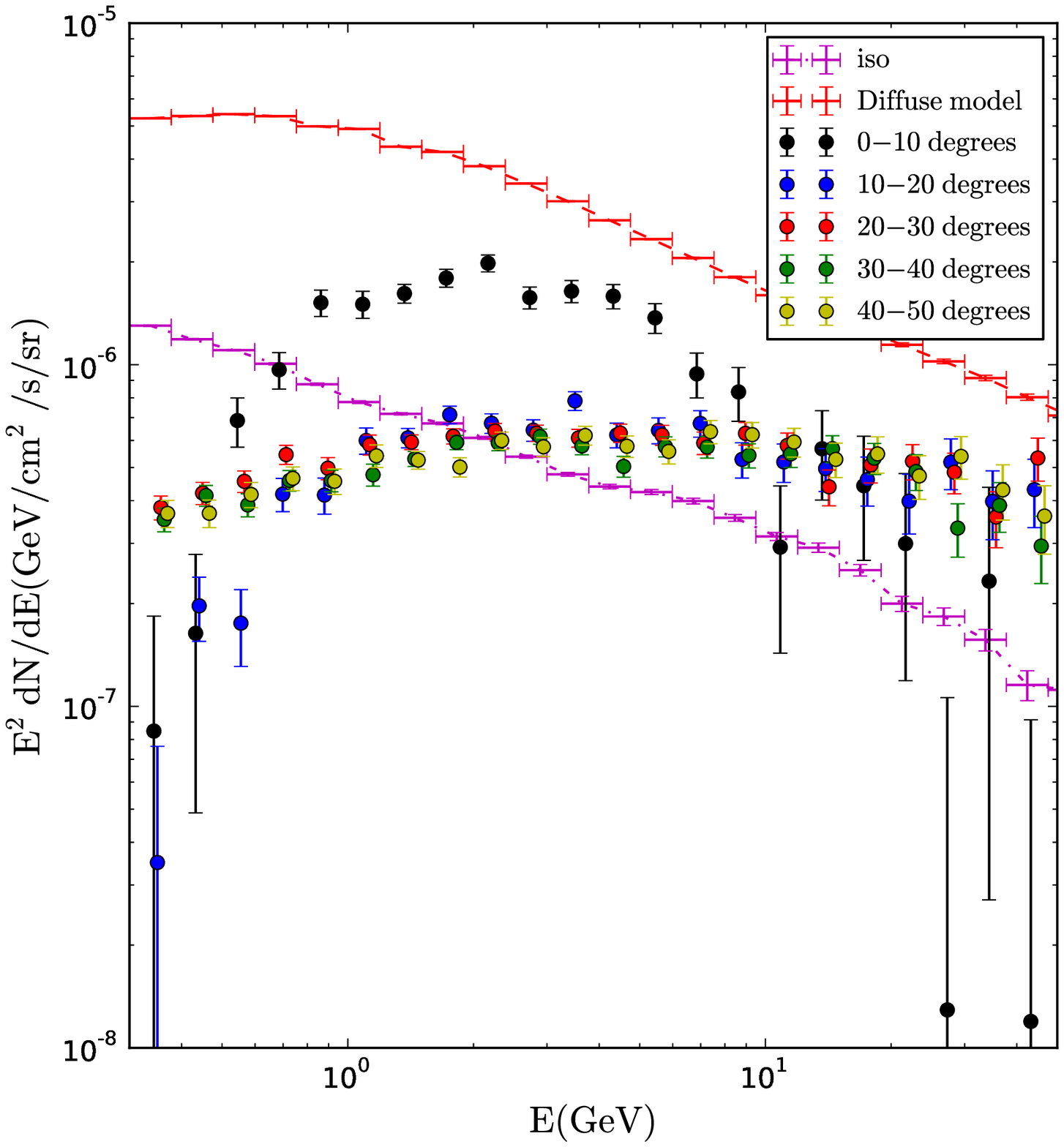}
        \includegraphics[width=0.48\linewidth]{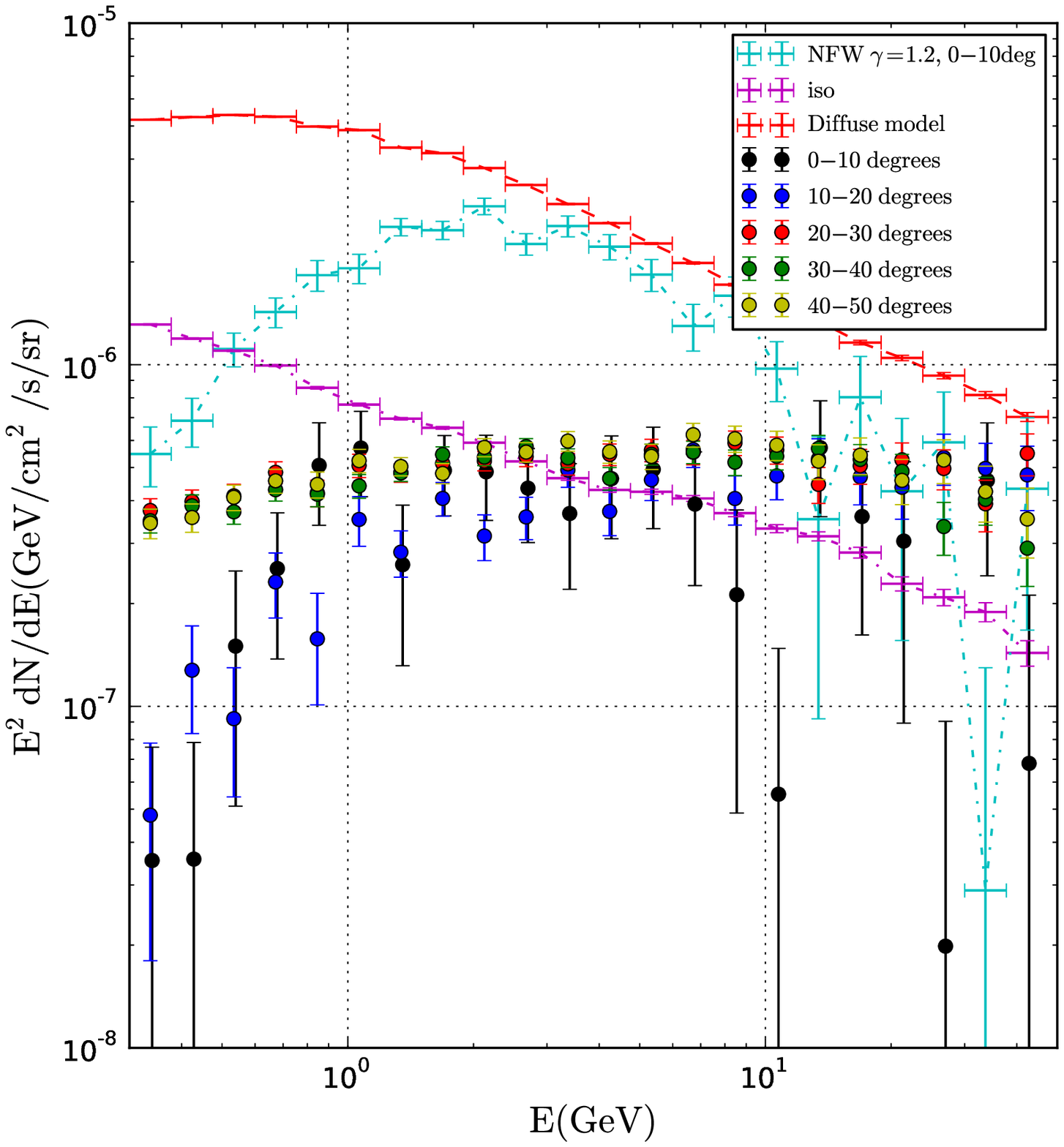}
    \end{center}
    \caption{The dark-matter-like GeV excess identified in our analysis of the Fermi Bubble regions with the P6V11 Galactic diffuse template. Left: The SED of five bubble slices and the corresponding background templates. Right: the same as the left panel except that the dark matter template is added.  Our results are consistent with that found in \cite{Hooper:2013rwa}, i.e., in both the slices of $0^{\circ}<b<10^{\circ}$ and $10^{\circ}<b<20^{\circ}$, a dark-matter-like GeV excess component is highly preferred.}
\label{fig:p6v11}
\end{figure}

We make use of Fermi Science Tools (v9r32p5) to complete data selection, calculate the exposure maps and convolve model templates with the PSF. For everything else we use our own code.
As a test of our code, we firstly analyze the data with P6V11 Galactic diffuse emission template that has been widely used in the GeV excess analysis. Following \cite{Hooper:2013rwa} we have analyzed the possible GeV excess in the Fermi Bubble regions, which have been divided into five slices. The results are presented in Fig.~\ref{fig:p6v11}. Unaccounted gamma ray emission presents in the low latitude regions ($b<20^\circ$).
Intriguingly, after the incorporation of the dark matter template, these ``unexpected" gamma-rays disappeared and the resulting spectra in all the five Bubble slices are almost the same. These results are remarkably consistent with that of \cite{Hooper:2013rwa} and in turn suggest that our code is reliable.
In the right panel of Fig.~\ref{fig:p6v11}, at energies of several hundreds MeV the spectra of the Bubble slices within $0^\circ-10^\circ$ and $10^\circ-20^\circ$ are different from the three high-latitude slices.
The reason is likely that at such low energies the angular resolution of Fermi-LAT is not good and hence there is strong coupling between the Bubble radiation and the dark matter template in low latitudes. Above 10 GeV, there is some coupling between the excess component and the Bubble radiation in the first slice because of the limited photon statistics.
Data smoothing could suppress the coupling and then make features more distinct. Whereas in our fit the whole Bubble instead of five slices are used. This saves some computational time, and could also avoid the coupling between the Fermi Bubble and the dark matter template. Then we fit the sky-map with each of the 128 groups of galactic diffuse emission templates and other four templates, in some ROIs defined in Sec.\ref{Sec:Templates}.

The fits are performed by combining all energy bins together. For the three sub-components of galactic diffuse emission (i.e.,  the $\pi_{0}$, Bremsstrahlung and IC templates) and the dark matter component we take into account the spectral information and there are three free parameters for the DGE component and one free parameter for the NFW component. Note that the spectra of $\pi_0$ and Bremsstrahlung are different, despite the similarities in their morphologies. That is why we do not treat them together.  For the Bubble and isotropic diffuse emission, we do not know the spectra and simply leave them free.

\subsection{Significance of the additional $b\bar{b}-$like excess component in different DGE models}\label{sec:IIIB}

\begin{figure}
    \begin{center}
        \includegraphics[width=0.48\linewidth]{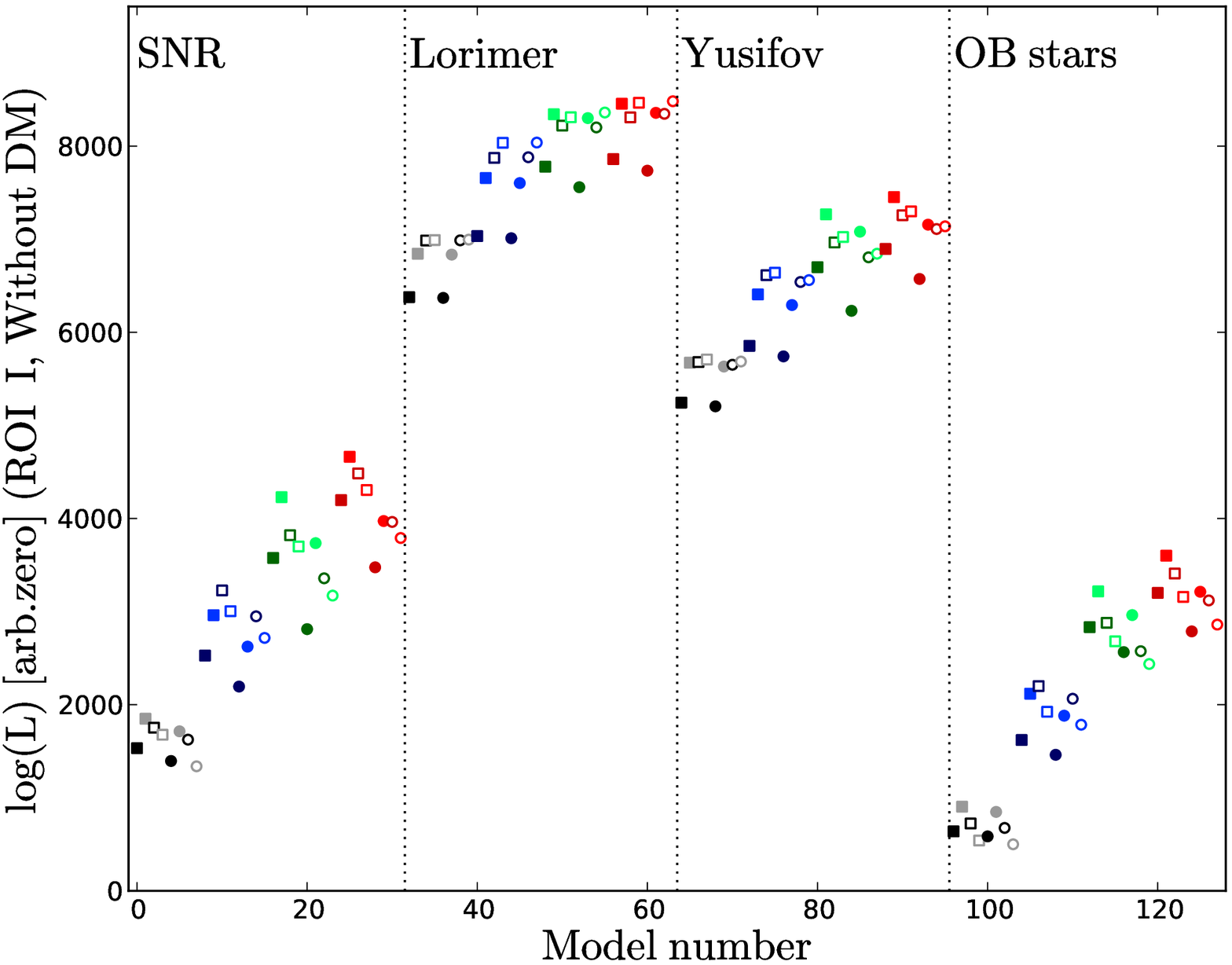}
        \includegraphics[width=0.48\linewidth]{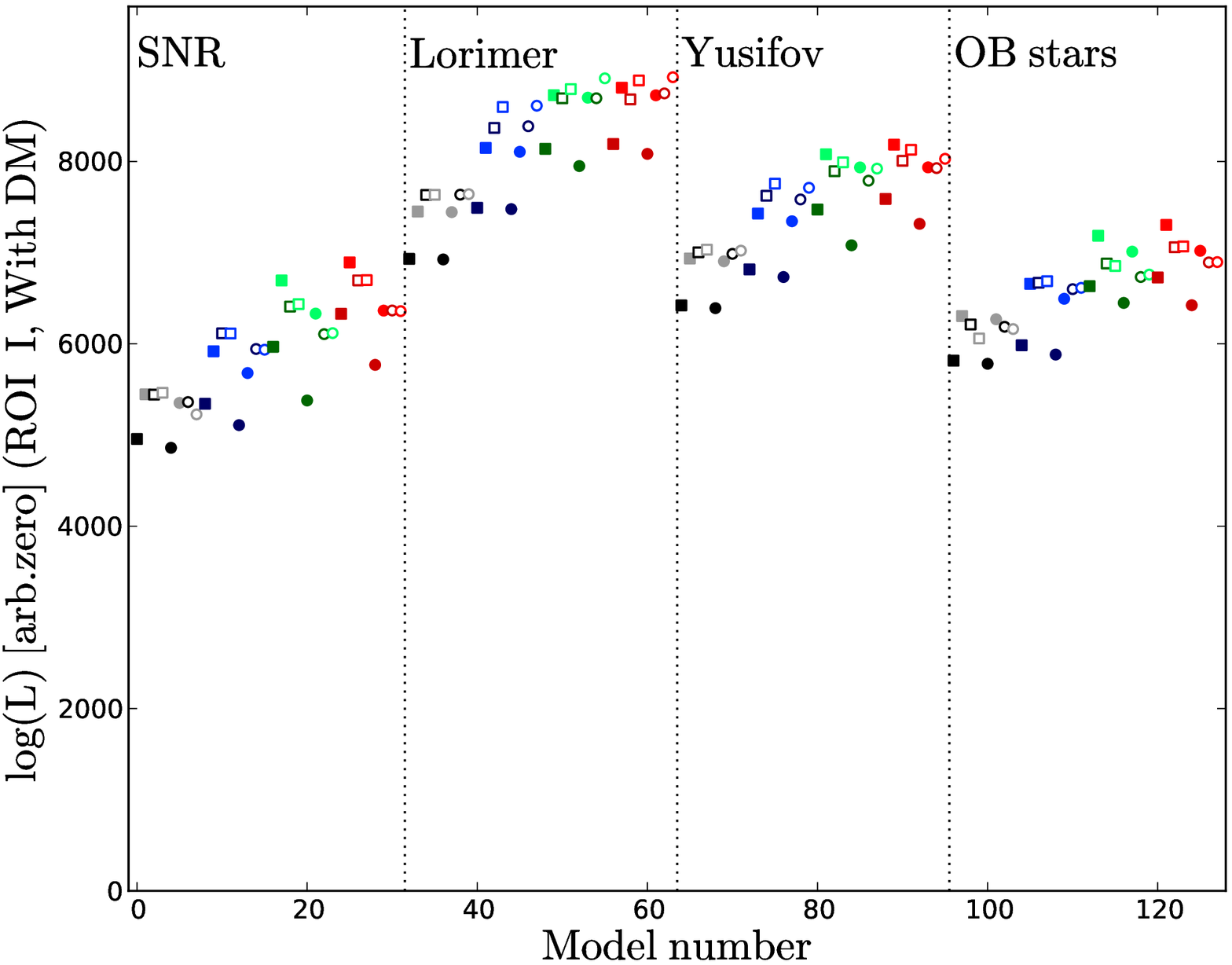}
        \includegraphics[width=0.48\linewidth]{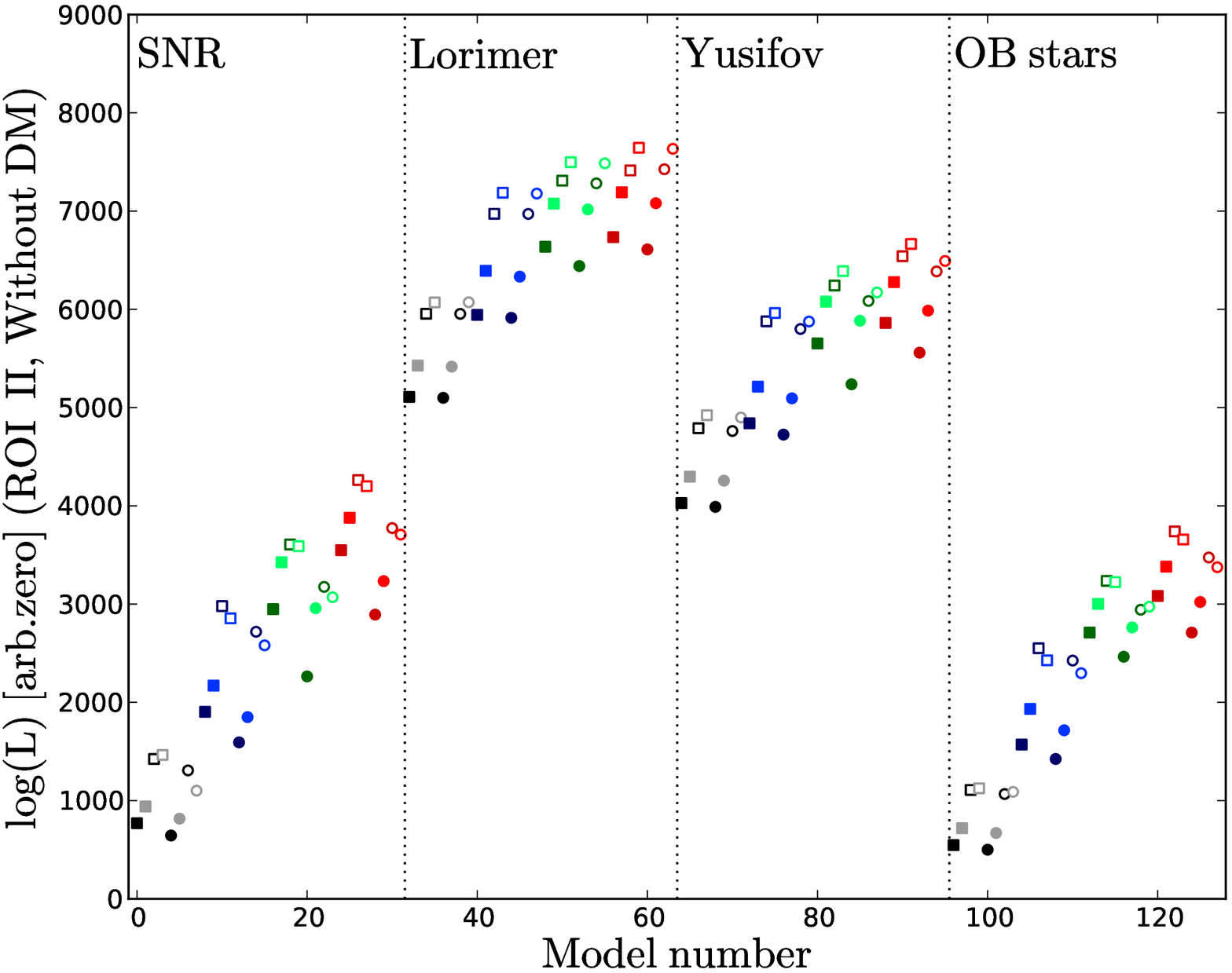}
        \includegraphics[width=0.48\linewidth]{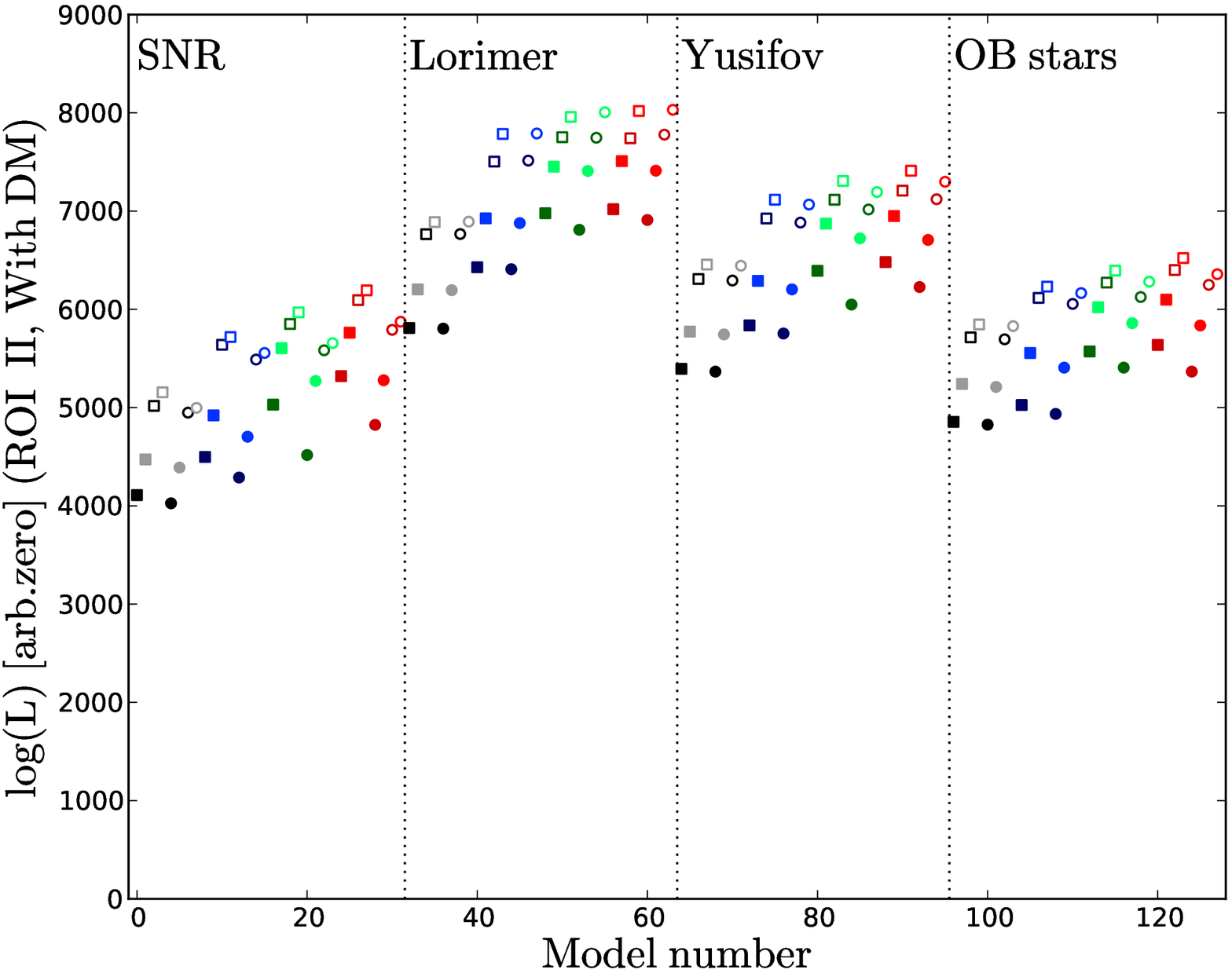}
        \includegraphics[width=0.48\linewidth]{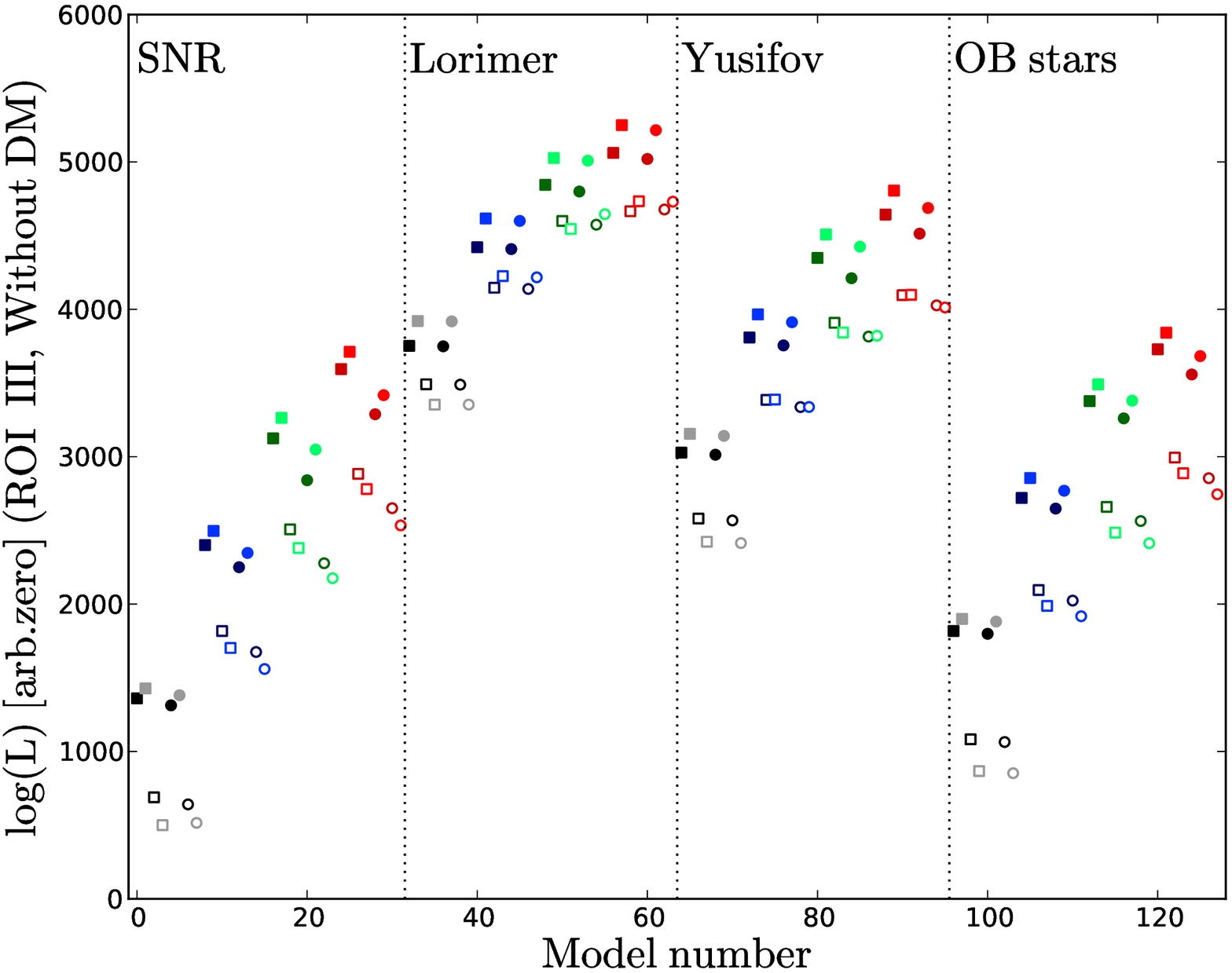}
        \includegraphics[width=0.48\linewidth]{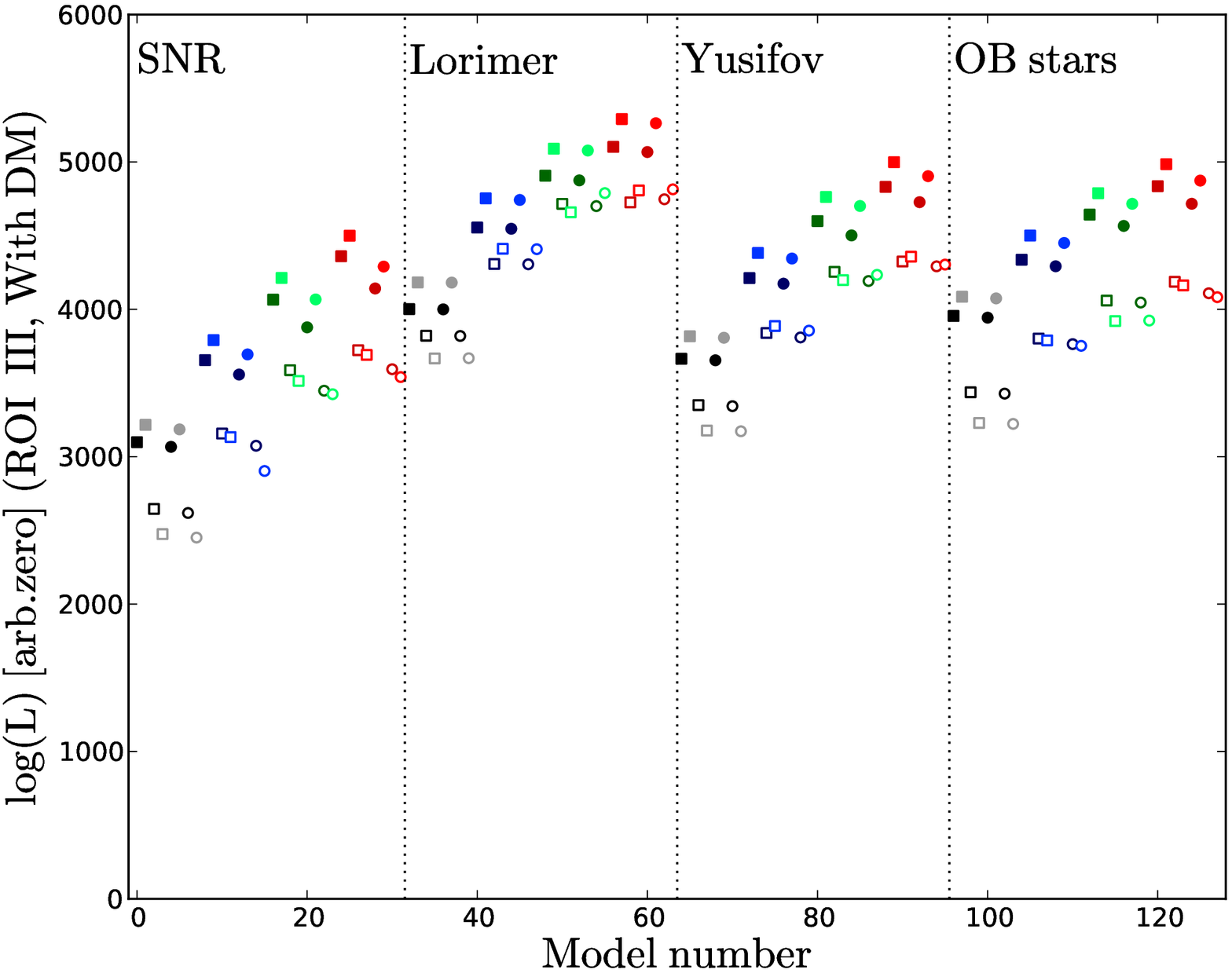}
    \end{center}
    \caption{Log-likelihood values obtained in the separate fits for each group of galactic diffuse gamma ray emission template: without and with the dark matter component (the left and right panel, respectively).
        The zero levels of the log-likelihood values are arbitrary but are equivalent in the same region. Therefore for a given region, the value difference between two models represents their likelihood ratio. For each group of template, three regions of interest defined in Fig.\ref{fig:masked} are considered, including  ROI I for the top panel, ROI II for the middle panel and ROI III for the bottom panel.
         In each panel, the model numbers as well as the color settings are the same as that used in \cite{FermiLAT:2012aa}: $z_{\rm h}=(4,~6,~8,~10)$ kpc are in (black, blue, green, red), respectively; $R_{\rm h}=(20,~30)$ kpc are represented by squares and circles, respectively; the filled and open points are for $T_{\rm s}=(150~{\rm K}~,{10^5~\mathrm{K}})$, respectively; and the dark and light colors are for $E(B-V)=(2,~5)$ magnitude cuts, respectively. The dotted vertical lines are the boundary of different CR source distribution models.
}
\label{fig:lnlh}
\end{figure}

The main goal in this subsection is to investigate whether a GeV excess likely originated from annihilation of dark matter particles to $b\bar{b}$, as found in \cite{Daylan:2014rsa}, is still needed in different DGE models or not.
For such a purpose we proceed to analyze the possible GeV excess with the 128 DGE templates, respectively. The main results, i.e., the log-likelihood values of these fits in the ROIs, are presented in Fig.~\ref{fig:lnlh}, where the left and right columns are in the cases of without and with the dark-matter-like excess component, respectively. The addition of a dark-matter-like radiation component indeed improves the goodness of all these fits. As usual, some DGE templates work better than the others. For example, for the DGE templates incorporating the pulsar-traced cosmic ray distribution model, the fits to the data yield larger log-likelihood values than those incorporating either the OB star-traced cosmic ray distribution model \cite{Bronfman:2000tw} or SNR-traced cosmic ray distribution model \cite{Case:1998qg}. Moreover, the fits with the DGE templates incorporating  Lorimer's pulsar-traced cosmic ray distribution model \cite{Lorimer:2006qs}, in which the source spatial distribution is set to be zero at $R =0$, yield larger log-likelihood value than those incorporating Yusifov's pulsar distribution model \cite{Yusifov:2004fr}. Interestingly, in the fits to the nuclei data the lowest $\chi^{2}$ was also obtained in Lorimer's pulsar-traced cosmic ray distribution model \cite{Lorimer:2006qs} (see the left panel of Fig.35 in \cite{FermiLAT:2012aa}).

A general trend shown in Fig.~\ref{fig:lnlh} is that the higher the $z_{\rm h}$ the larger the likelihood, consistent with that found in the cosmic ray modeling in \cite{FermiLAT:2012aa}. No strong dependence of the log-likelihood values on $R_{\rm h}$ is found in Fig.~\ref{fig:lnlh}.
However, the trend of $T_{\rm s}$ is different between the fitting regions and is correlated with the $E(B-V)$ cut. For modeling performed in ROI I, when a $E(B-V)$ cut of 2 mag is adopted, all the fits prefer to a larger $T_{\rm s}$. While there is no constant favored $T_{\rm s}$ value when $E(B-V)$ cut of 5 mag is assumed. This trend shows no bias between different $R_{\rm h}$ values and different source distributions. In ROI II, it shows a simpler trend: all the fit favor a greater $T_{\rm s}$ value. What's more, the difference between the log-likelihood values is the largest in the case of $z_{\rm h}$ = 6\,kpc. An interesting phenomenon happens when a larger galactic plane has been masked (i.e., in ROI III). All the fits favor a smaller $T_{\rm s}$ value, which is contrary to that found in ROI II.
$E(B-V)$ also plays a role in modifying the likelihood value, in particular if the dark matter component has been excluded. For instance, in ROI I  all fits favor a $E(B-V)$ cut of 5 magnitudes in the two kinds of pulsar-traced cosmic ray distribution models. In the SNR-traced and OB star-traced cosmic ray distribution models, for $T_{\rm s}=150$ K the fits also favor the $E(B-V)$ cut of 5 magnitudes, while in the optical thin scenarios (i.e., $T_{\rm s}=10^5~\mathrm{K}$) the fits favor the 2 magnitude cut of $E(B-V)$.

\begin{figure}
  \begin{center}
    \includegraphics[width=0.48\linewidth]{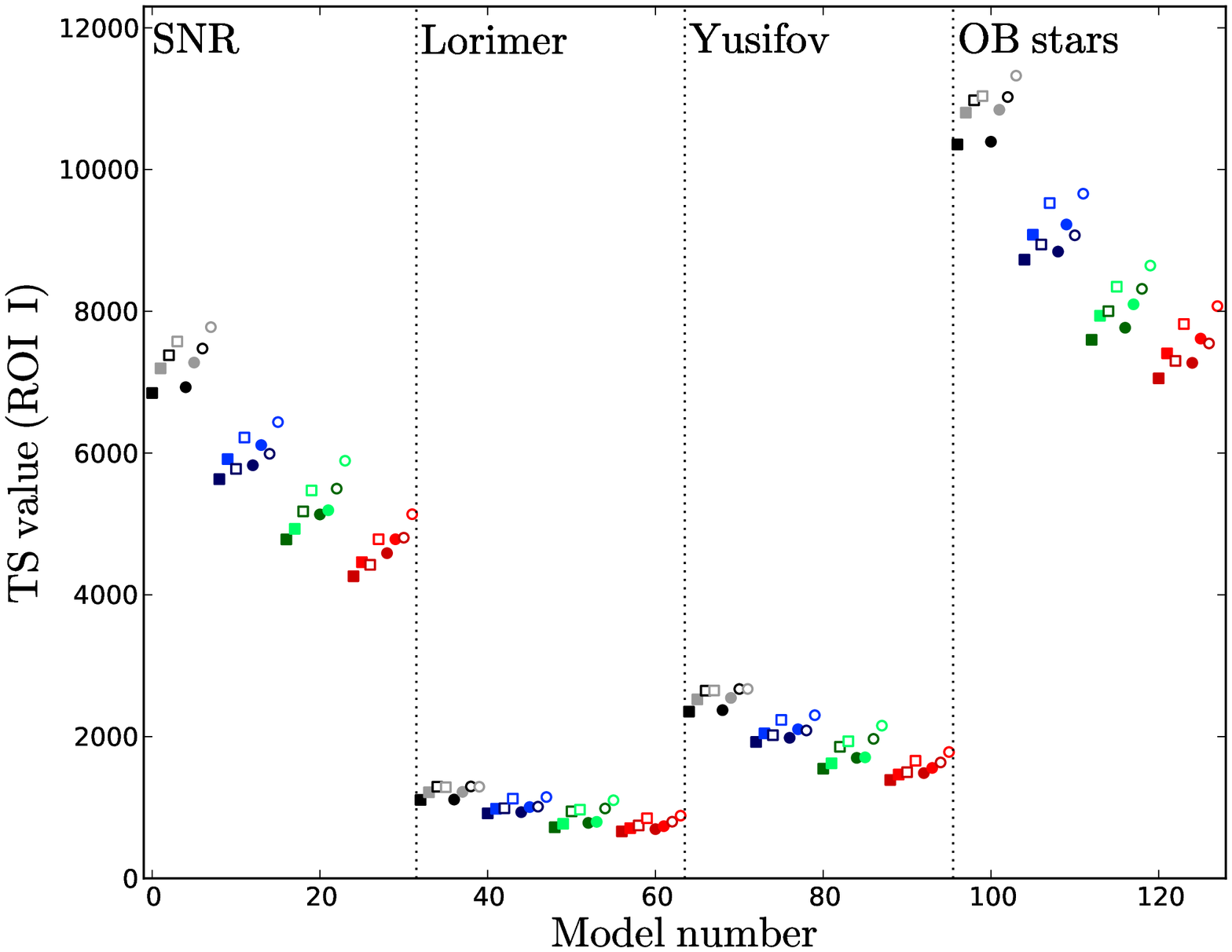}\\
    \includegraphics[width=0.48\linewidth]{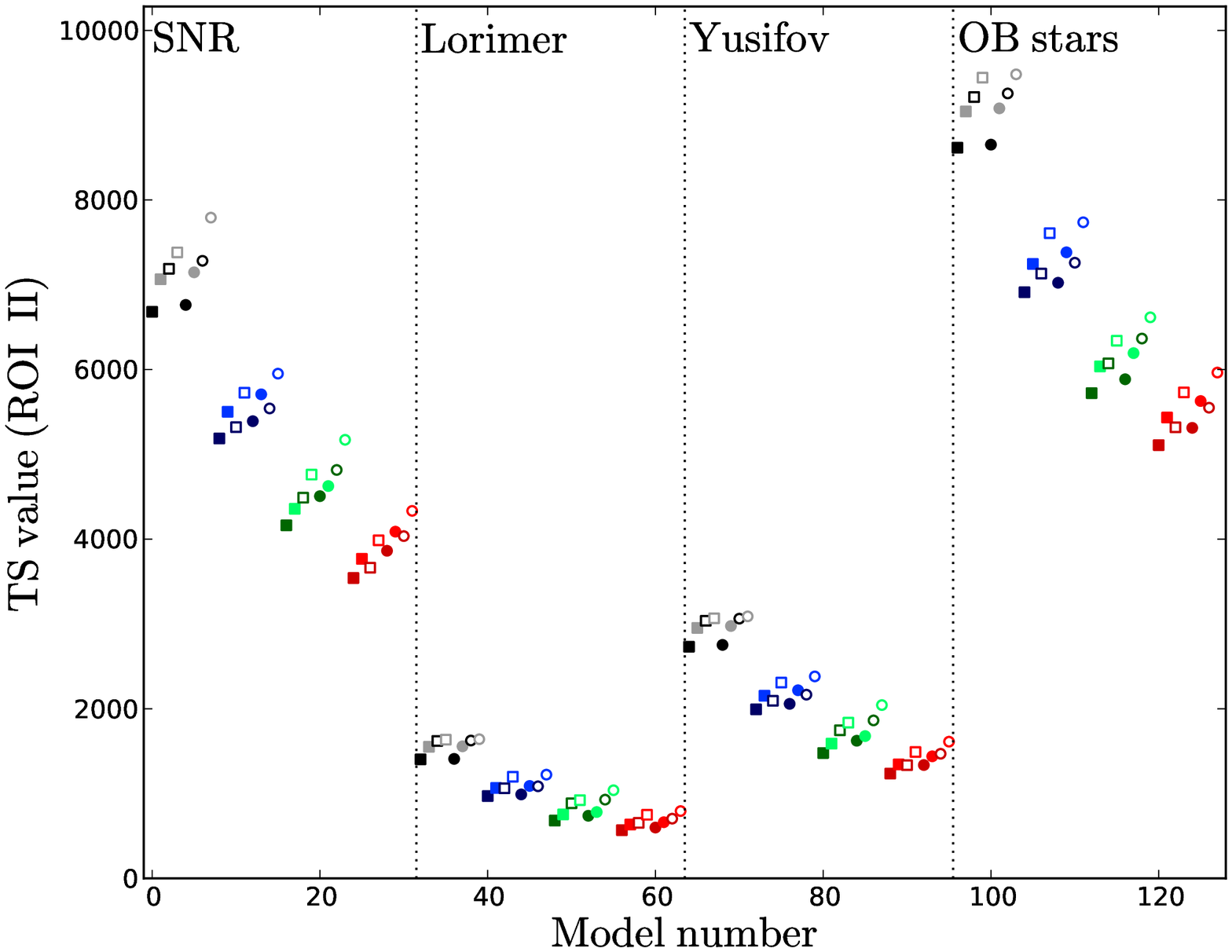}
    \includegraphics[width=0.48\linewidth]{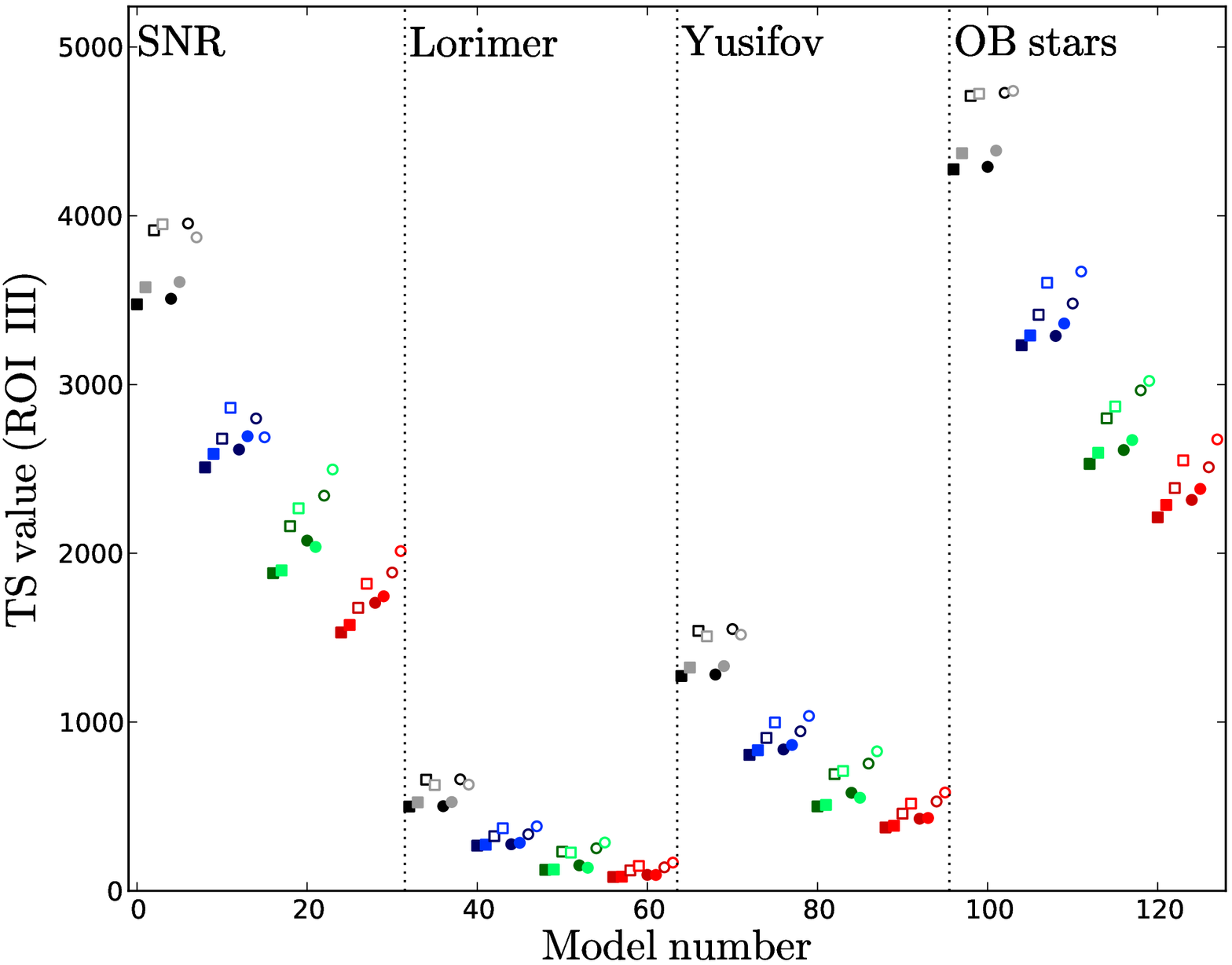}
  \end{center}
    \caption{TS values for the presence of an additional dark matter-like component obtained in the fits with different galactic diffuse emission templates
    (See Fig.~\ref{fig:lnlh} for legend).
}
\label{fig:ts}
\end{figure}

\begin{figure}
  \begin{center}
    \includegraphics[width=0.48\linewidth]{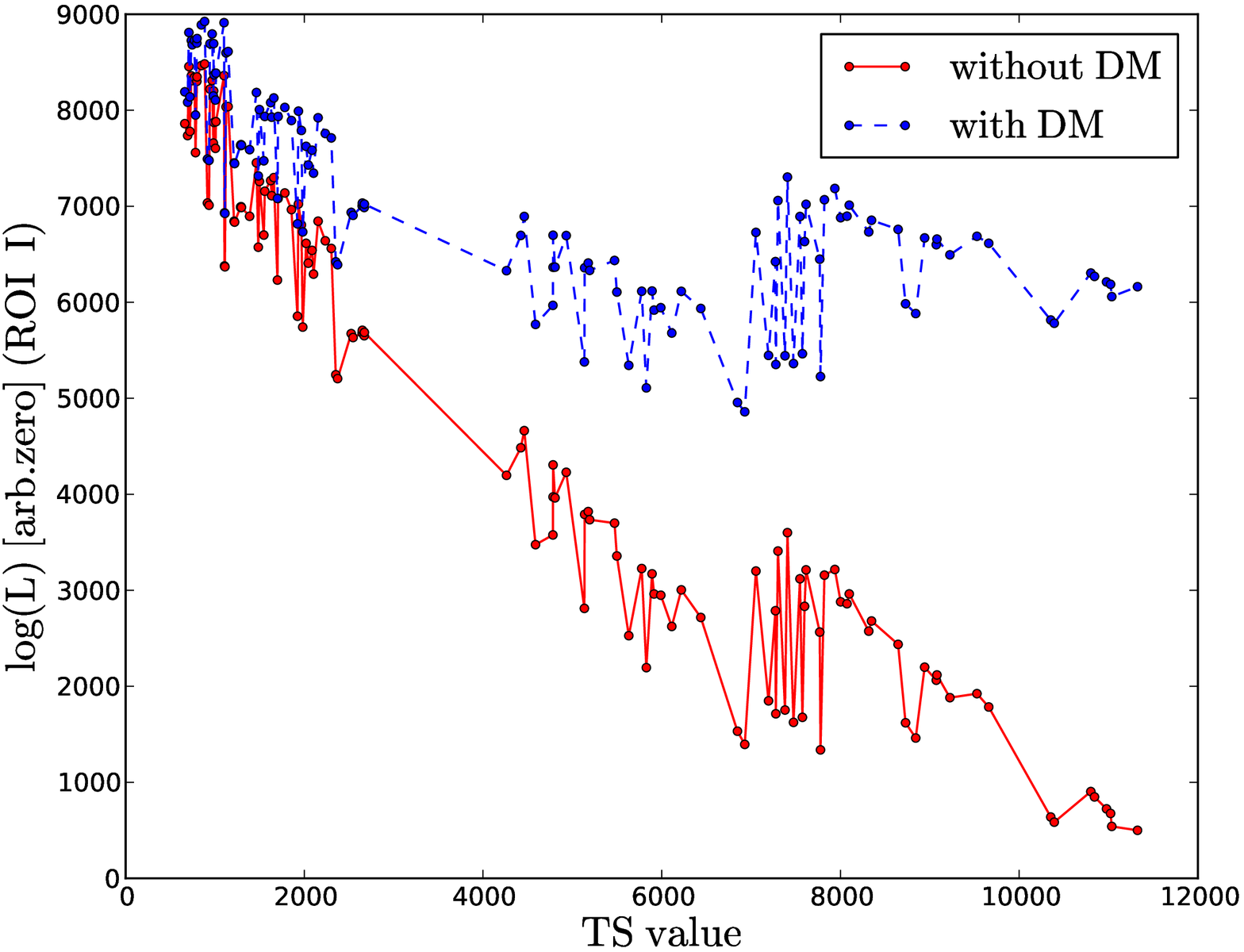}\\
    \includegraphics[width=0.48\linewidth]{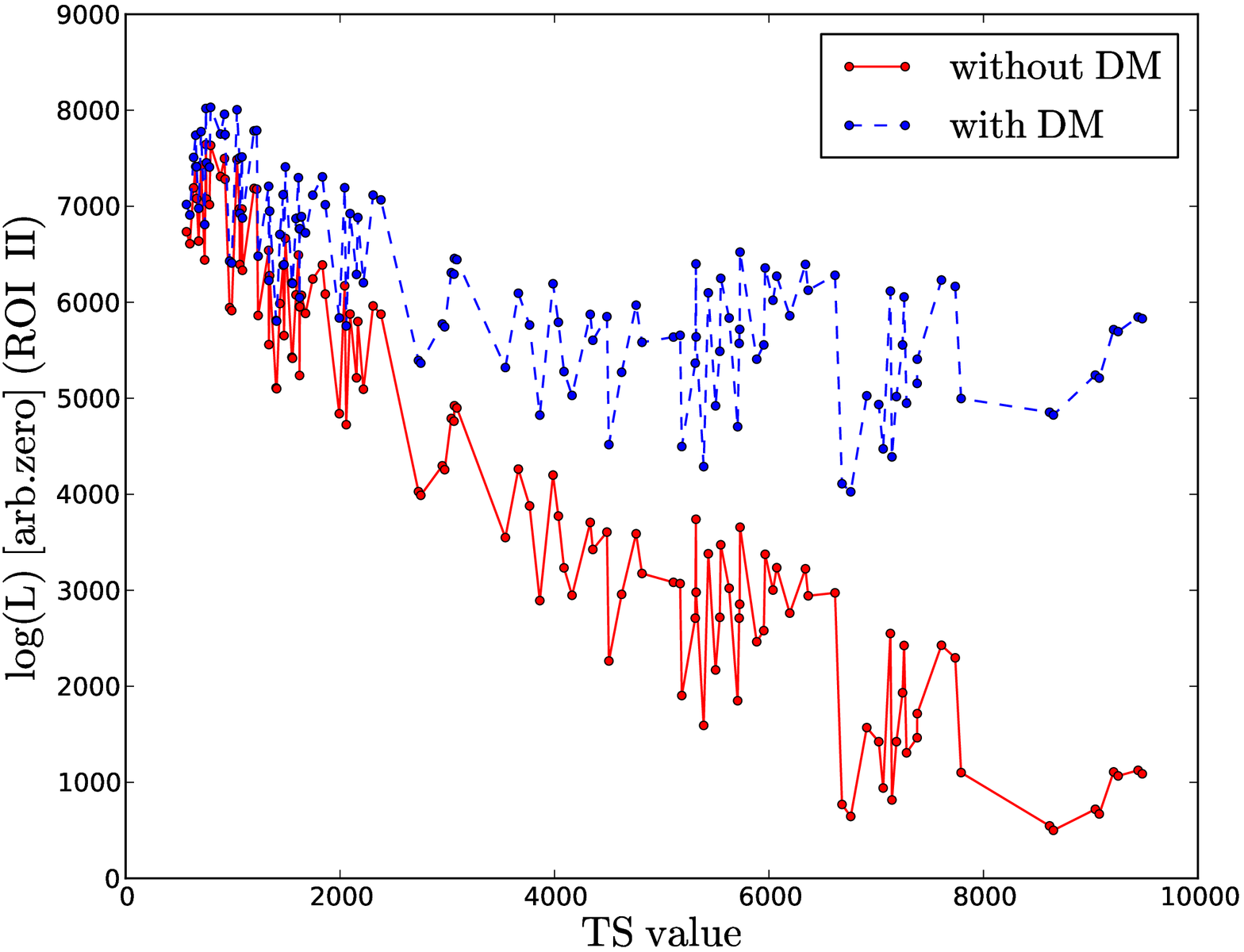}
    \includegraphics[width=0.48\linewidth]{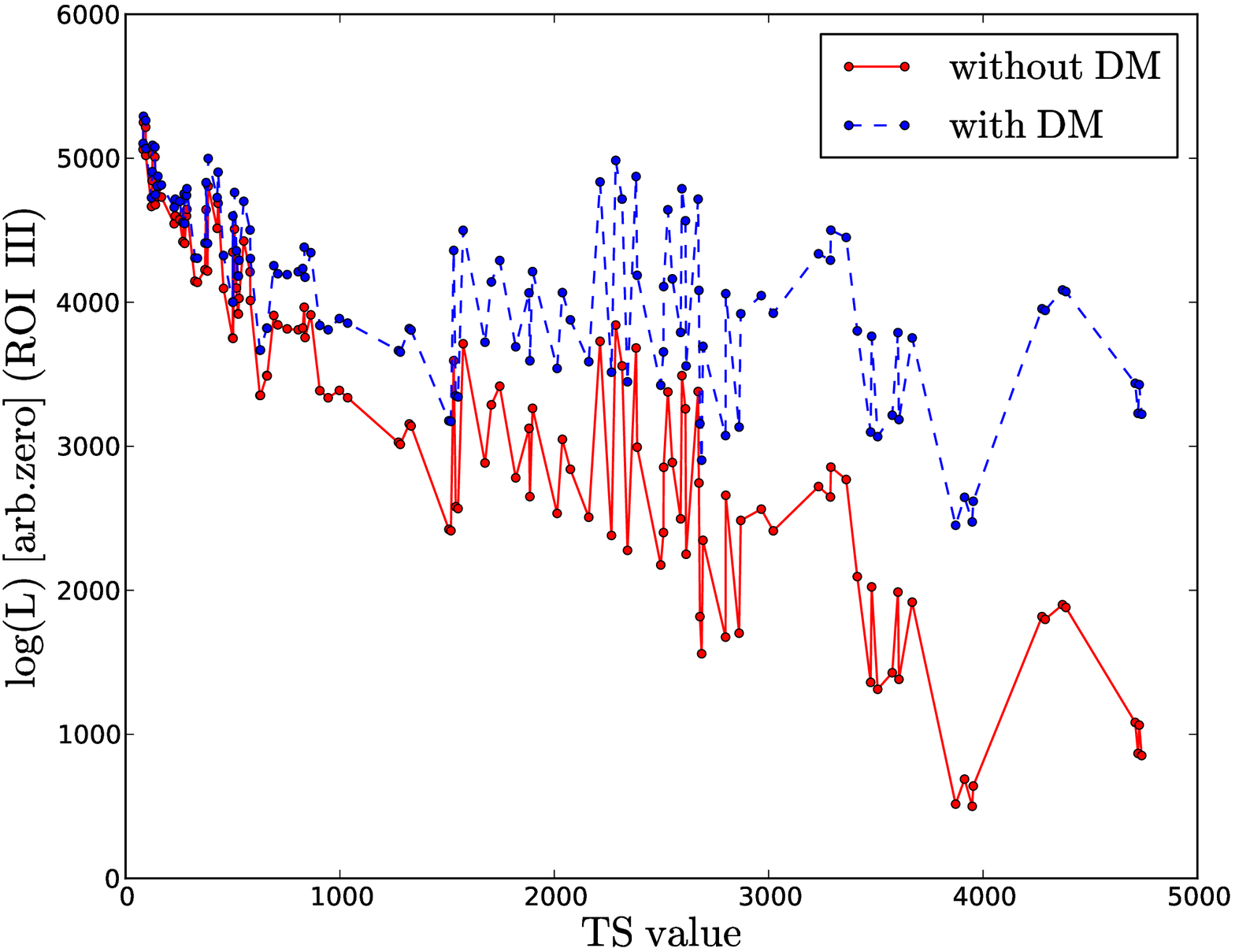}
  \end{center}
    \caption{Log-likelihood (adopted from Fig.~\ref{fig:lnlh}) versus TS values (adopted from Fig.~\ref{fig:ts}) in three regions of interest defined in Fig.~\ref{fig:masked}. The solid (dashed) lines are the log-likelihood values without (with) an additional dark-matter-like component.
}
\label{fig:lhts}
\end{figure}

In Fig.~\ref{fig:ts} we present the TS values for the presence of an additional dark matter-like component obtained in the fits with different galactic diffuse emission templates. Such TS values are straightforwardly obtained by directly subtracting the values of the right panels from the values of the corresponding left panels of Fig.~\ref{fig:lnlh} and then multiplying by a factor of $-2$. The TS values of the GeV excess displayed in ROI I and ROI II are much larger than the corresponding ones in ROI III,
simply due to the much lower signal-to-noise in the relatively high latitude region.
Intriguingly, after masking the whole Fermi Bubbles (which extends to $b = 0^\circ$ in the north and $b = -5^\circ$ in the south) the TS value does not decrease considerably, suggesting that the GeV excess component is intrinsic and is not part of the Fermi bubble radiation. The minimal TS value we find for the additional dark-matter-like excess component is $\approx 670$ in the region of $|b|>5^\circ$ and $\approx 82$ in the region of $|b|>10^\circ$, suggesting that the excess is indeed statistically significant. The corresponding velocity-averaged cross section is $\langle \sigma v\rangle\sim 0.6-2\times 10^{-26}~{\rm cm^{3}~s^{-1}}~(\rho_0/0.43~{\rm GeV~cm^{-3}})^{-2}$, where $\rho_0$ is the local energy density of the dark matter. Some interesting trends of TS values on the input parameters are evident in Fig.~\ref{fig:ts}, too. The most remarkable one may be that the Lorimer's pulsar-traced cosmic ray distribution models have the lowest TS values. This is reasonable since such a kind of cosmic ray distribution model has the lowest $\chi^{2}$ in modeling the nuclei data \cite{FermiLAT:2012aa} and the largest log-likelihood value in our gamma ray fitting (see Fig.~\ref{fig:lnlh}), i.e., the difference between the data and the background templates is the smallest. For the same reason, the templates with larger log-likelihood values tend to have smaller TS values (see Fig.~\ref{fig:lhts}).

\subsection{The spectrum-energy distribution (SED) of the additional NFW profile-like component in different DGE models}\label{sec:IIIC}

\begin{figure}
    \begin{center}
        \includegraphics[width=0.58\linewidth]{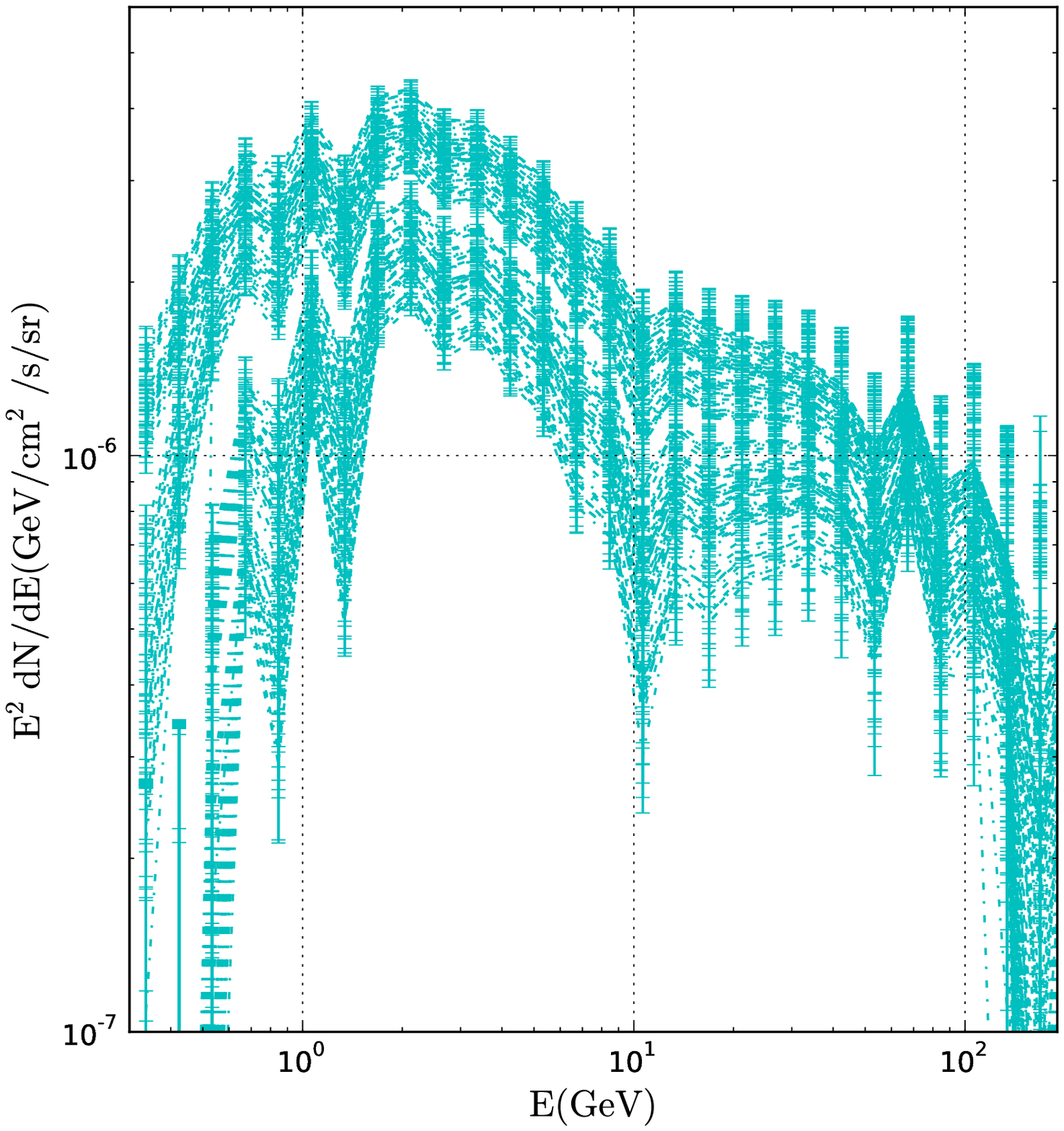}
    \end{center}
    \caption{The spectrum energy distribution of the GeV excess, averaged within the $10^{\circ}$ from the galactic center and assuming a generalized NFW profile with an inner slope $\gamma$ = 1.2, for all the 128 GDE models. Note that the fits are preformed in ROI I.  The excess peaking in the energy range of $1-3$ GeV is distinct in all fits.}
\label{fig:DMseds}
\end{figure}

\begin{figure}
    \begin{center}
        \includegraphics[width=0.78\linewidth]{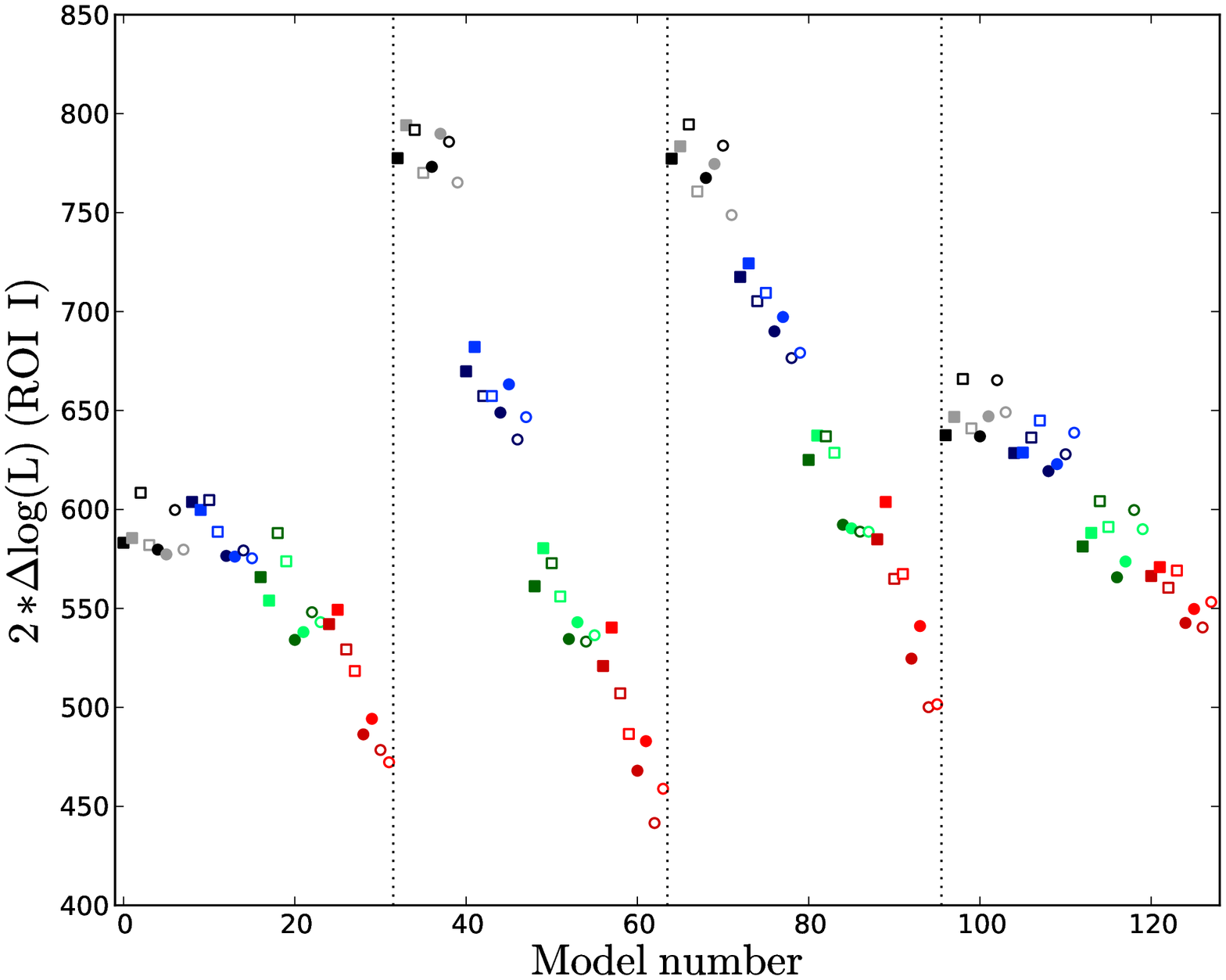}
    \end{center}
    \caption{The further increase of the 2*log(Likelihood) values of the fits to the data with an additional spatially-extended component thanks to the relaxing of the SED.}
    \label{fig:ts_dmseds}
\end{figure}

In the last subsection, the spectrum of the GeV excess has been fixed to be that of the gamma-rays originating from the annihilation of $\sim 35$ GeV dark matter particles into $b\bar{b}$. In this subsection such a strong restriction is relaxed and we examine the role of the DGE models in shaping the SED of the `potential' excess component. Note that the treatments are the same as in Sec.\ref{sec:IIIB} except that the spectrum-energy distribution of the `potential' excess component is not fixed any longer.
For simplicity, the global fits are performed to the data in ROI I. The resulting SEDs of the additional component are presented in Fig.~\ref{fig:DMseds}. Comparing with the fits performed with a fixed SED in Sec.\ref{sec:IIIB}, the goodness of current fits have been considerably improved (see Fig.\ref{fig:ts_dmseds}), as expected. In Fig.~\ref{fig:DMseds} there are two remarkable features. One is that the SED in each spectral fit has a distinct peak at energies of $1-3$ GeV, which suggests that the GeV excess is indeed intrinsic, strengthening the conclusion made in Sec. \ref{sec:IIIB}. The other is the presence of a non-ignorable high-energy ($>10$ GeV) component in the SEDs, which is in agreement with that found by Calore et al. \cite{Calore:2014xka} and Murgia \cite{Murgia}.

\begin{figure}
    \begin{center}
        \includegraphics[width=0.78\linewidth]{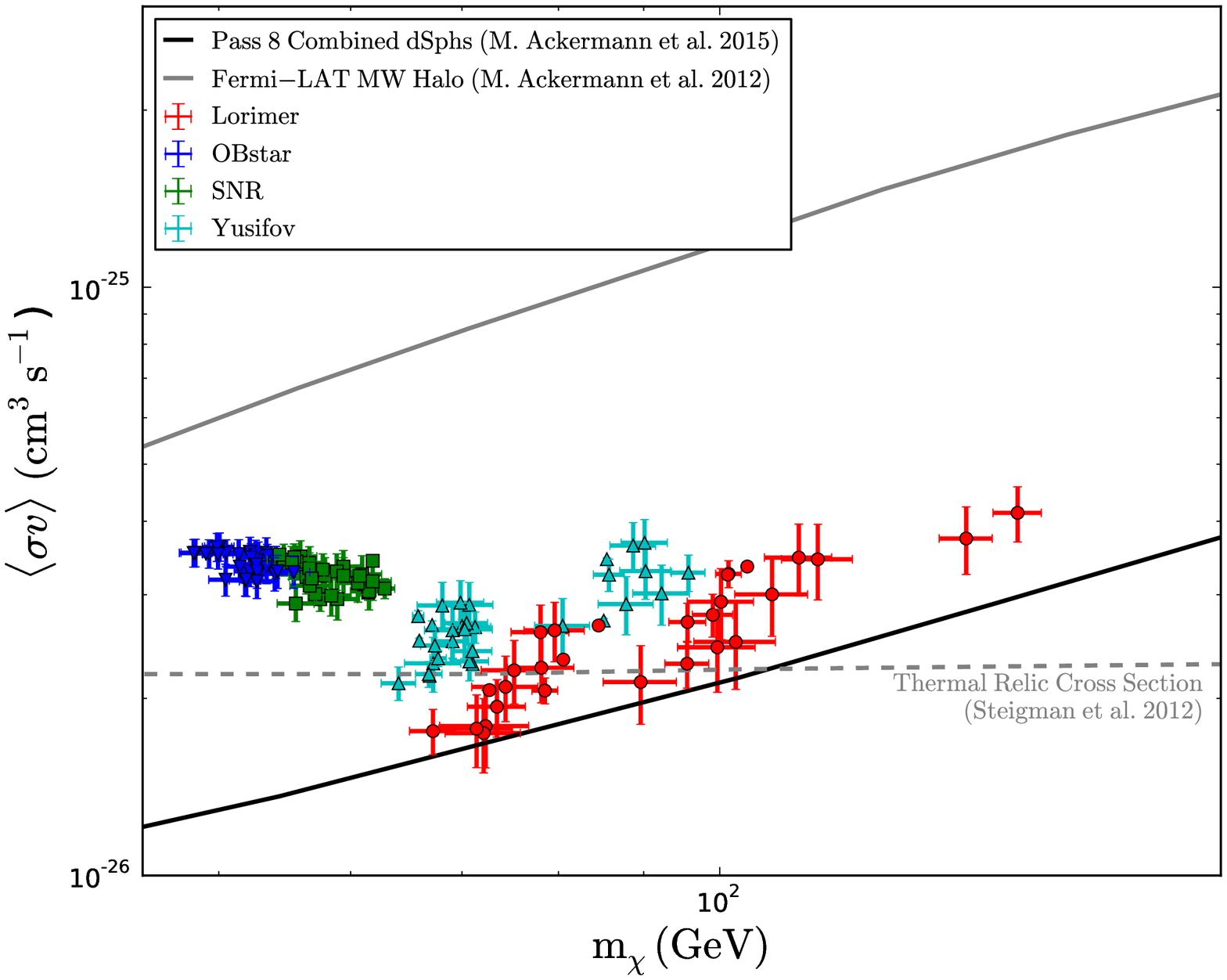}
    \end{center}
    \caption{The $\langle \sigma v \rangle$ and $m_{\chi}$ obtained in the $b\bar{b}$-annihilation-channel spectral fits to the GeV excess SED obtained in different DGE models. Note that the red, blue, green and cyan represent the four kinds of the cosmic-ray distribution models (i.e., 'Lorimer', 'OBstar', 'SNR' and 'Yusifov'), respectively. The dashed line is the latest thermal relic cross section of dark matter annihilation \cite{Steigman:2012nb}. We also show constraints for dark matter from Milky Way halo \cite{Ackermann:2012rg}  and dwarf galaxies \cite{Ackermann:2015zua}. }
\label{fig:bbar}
\end{figure}

In Fig.~\ref{fig:bbar} we fit the resulting SEDs with the $\gamma-$ray spectrum originated from annihilation of dark matter particles into $b\bar{b}$. 
The local energy density of the dark matter particles is assumed to
be $0.43~{\rm GeV~cm^{-3}}$. Not surprisingly, the velocity-averaged cross sections of dark matter particle annihilation in most fits are found to be  in tension with the tight constraints reported in the literature (see Fig.~\ref{fig:bbar}). Nevertheless, in the Lorimer's pulsar-traced cosmic ray distribution model, quite a few fits yield $\langle \sigma v \rangle$ that is still consistent with the strictest limit set by the latest Pass 8 data analysis of some dwarf galaxies \cite{Ackermann:2015zua}. Hence the dark matter origin model for the Galactic GeV excess has not been ruled out, yet.

\section{Discussion}
In this work we have analyzed the gamma-ray emission measured by the Fermi Gamma-ray Space Telescope from the inner regions of the Milky Way. In total 128 Galactic diffuse emission background templates/models have been taken into account, in each of them the possible dark-matter-originated radiation (i.e., annihilating to $b\bar{b}$) components in the regions of $l<80^\circ$ and $|b|>5^\circ$ (including and excluding the Fermi Bubbles) or $|b|>10^\circ$ have been explored and the dark matter energy density distribution has been taken as the generalized NFW profile with $r_{\rm s}=20$ kpc and the slope index $\alpha=1.2$.  The minimal TS value we find for the dark-matter-like excess component is $\approx 670$ in the region of $l<80^\circ$ and $|b|>5^\circ$ and $\approx 82$ in the region of $l<80^\circ$ and $|b| >10^\circ$ (see Sec.\ref{sec:IIIB}), strongly suggesting that the excess is indeed statistically significant and robust. The corresponding cross section of the dark mater particles with rest mass $m_\chi\sim 35$ GeV annihilating into $b\bar{b}$ is $\langle \sigma v\rangle\sim 0.6-2\times 10^{-26}~{\rm cm^{3}~s^{-1}}~(\rho_0/0.43~{\rm GeV~cm^{-3}})^{-2}$, consistent with what was found in \cite{Daylan:2014rsa}. Furthermore, we have examined the role of Galactic diffuse emission background templates/models in shaping the spectrum-energy distribution of the excess component. The treatments are the same as in Sec.\ref{sec:IIIB} except that the shape of the SED is not fixed any longer. Distinct GeV excess still displays in all the fits, strengthening our previous conclusion that the GeV excess is robust. At higher energies (i.e., $>10$ GeV), however, there are some significant radiation, consistent with the finding of \cite{Calore:2014xka, Murgia} (see Sec.\ref{sec:IIIC}).

The presence of a spatially-extending GeV excess component that is well consistent with the signal expected in dark matter particle annihilation is very attractive. We caution that
some astrophysical objects may also give rise to rather similar GeV radiation signal. For example, a population ($\sim 10^{4}$) of less-luminous millisecond pulsars (MSPs) may be able to account for the GeV excess for the following reasons: (1) MSPs are known strong GeV gamma ray emitters that peak at a few GeV; (2) Estimates of the spatial distribution of M31 low mass X-ray binary population indicate that the number of MSPs located in the Galactic center  could scale
as steeply as $1/r^{2.4}$ \cite{Abazajian:2012pn}; (3) a
population of hard ($\Gamma <1$) ``under-luminous" MSPs either endemic to the innermost region
or part of a larger nascent collection of hard MSPs that appears to be emerging in the second {\it Fermi} LAT Pulsar Catalogue, which could reproduce
the observed flux of the GeV excess \cite{Mirabal:2013rba, Yuan:2014rca}. Alternatively, the ``Galactic center excess"
has been argued to be explained by a recent cosmic-ray injection burst, with an age in the $10^{3}-10^{4}$ year range, while the
extended ``inner Galaxy excess" has been suggested to point to mega-year old  cosmic-ray injection \cite{Carlson:2014cwa}. Distinguishing between the dark matter model and the astrophysical model is not a trivial task. The most straightforward way to confirm the dark matter origin of the GeV excess, if it is, may be the detection of a rather similar component in the nearby dwarf galaxies. In a dedicated study, 4-year gamma ray observation results of 25 dwarf spheroidal satellite galaxies of the Milky Way have been reported and a combined analysis of 15 dwarf galaxies
under the assumption that the characteristics of the dark
matter particle are shared between the dwarfs has been carried out. No
globally significant excess was found for any of the spectral
models tested. The largest deviation from the null hypothesis occurs for
soft gamma ray spectra and can be fitted by dark matter in the mass range
from 10 to 25 GeV annihilating to $\bar{b}b$ with a cross section in the order of $10^{-26}~{\rm cm^{3}~s^{-1}}$ for a ${\rm TS}\sim 8.7$ \cite{Ackermann:2013yva}.
However, such an attractive weak signal has not been confirmed by the latest Pass 8 data analysis of the dwarf galaxies \cite{Ackermann:2015zua}. A signal comparable with the Galactic GeV excess was claimed in a search for $\gamma$-ray emission from the direction of the newly discovered dwarf galaxy Reticulum 2 \cite{Geringer-Sameth:2015lua} while the Fermi-LAT collaboration did not confirm \cite{Drlica-Wagner:2015xua}. Nevertheless, as shown in Fig.\ref{fig:bbar}, the $\langle \sigma v\rangle$ obtained in a few fits is still consistent with the current strictest limit reported in \cite{Ackermann:2015zua}. The situation is thus unclear and further studies are highly needed to pin down the physical origin of the GeV excess.

Recently, PANGU (the PAir-productioN Gamma-ray Unit), a small astrophysics mission optimized for spectro-imaging, timing and polarisation studies in gamma rays, in the still poorly explored energy band from 10 MeV to a few GeV, has been proposed \cite{Wu:2014tya}. PANGU has excellent angular resolution, which is about 1 degree at 100 MeV and 0.2 degree at 1 GeV, much smaller than that of Fermi-LAT in such an energy range. With the considerably improved PSF, PANGU will resolve and separate potential gamma-ray sources in the inner Galaxy and thus help reveal the nature of the GeV excess.

\acknowledgments We thank M. Su, C. Weniger, R. Z. Yang and Q. Yuan for valuable communications. We also thank T. Porter and A. Strong for the help about GALPROP. This work was supported in part by 973 Programme of China under grant 2013CB837000, by National Natural Science of China under grant 10925315, by China Postdoctoral Science Foundation under grant 2014M551680, and by the Foundation for
Distinguished Young Scholars of Jiangsu Province, China (No. BK2012047).  YZF is also supported by the 100
Talents programme of Chinese Academy of Sciences.

$^\ast$Corresponding authors (xyhuang@pmo.ac.cn, xiangli@pmo.ac.cn, yzfan@pmo.ac.cn).

\bibliographystyle{apsrev}
\bibliography{library}

\begin{appendix}

\section{The influence of sources in 2FGL on the fit results}
In all the fits in the main text we simply took the source maps as part of the model and 
the source maps were fixed.
Here we test the validity of such an approximated treatment. In our `standard' analysis performed in Sec.\ref{sec:IIIB}, the minimum TS values for the additional dark matter-like GeV excess component are 663 in ROI I, 568 in ROI II and 82 in ROI III, respectively. The DGE template incorporating with the Lorimer-traced cosmic ray distribution model and the physical parameters including $z_h$ = 10 kpc, $R_h$ = 20 kpc, $T_s$ = 150 K and a $E(B-V)$ cut of 2 mag are involved in such fits.
We re-do the fits with the same DGE template and physical parameters. The difference is that the source maps are excluded and just the 200 most luminous sources have been masked. The purpose is to see whether these point/extended sources can modify the fit results significantly or not.
The minimum TS values we obtained are 732 in ROI I, 655 in ROI II and 69 in ROI III, respectively. All are reasonably consistent with that found in the `standard' analysis performed in Sec.\ref{sec:IIIB}.

\begin{figure}
\includegraphics[width=85mm,height=80mm]{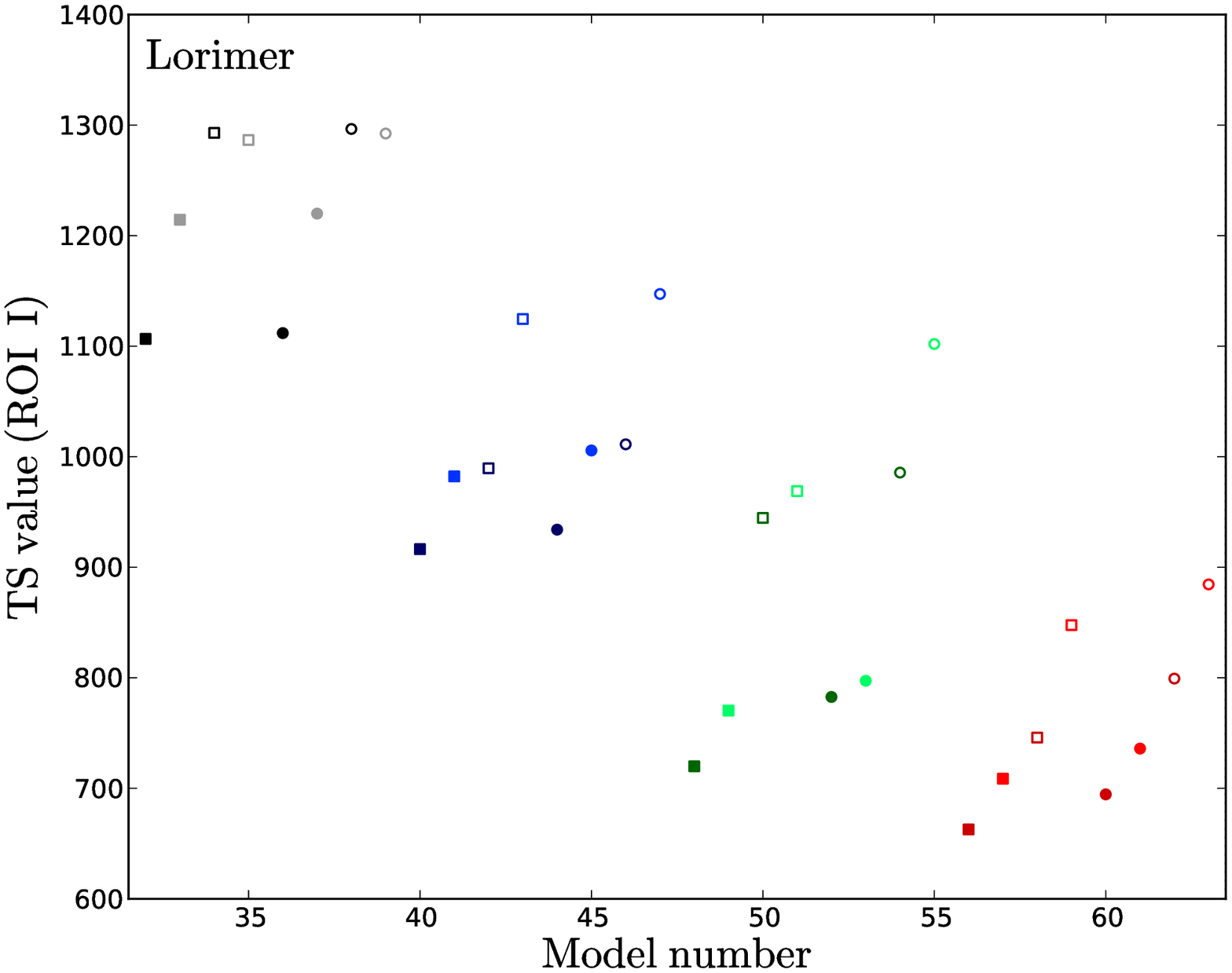}
\includegraphics[width=85mm,height=80mm]{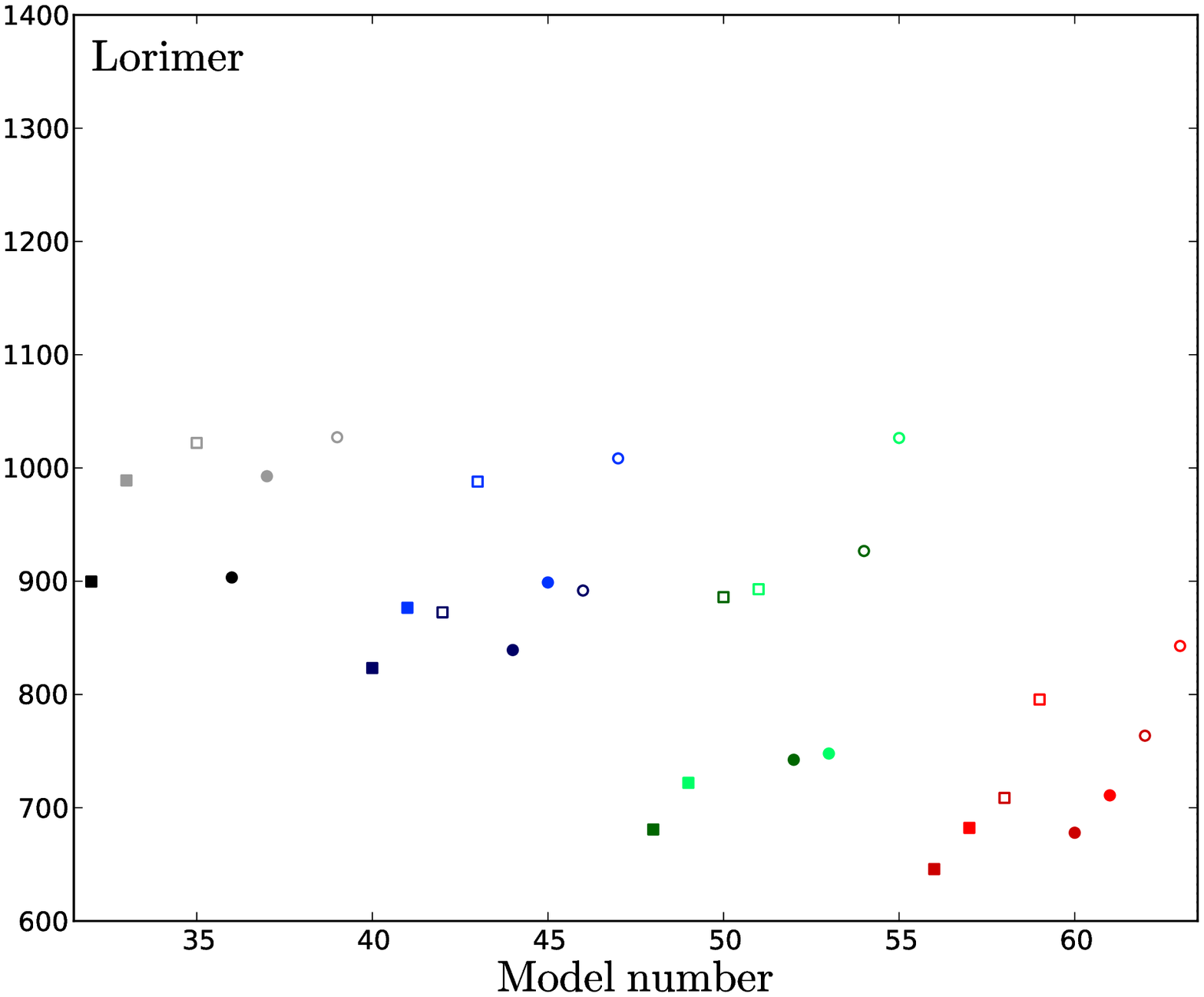}
\caption{The influence of sources in 2FGL on the fit results. The Lorimer's pulsar-traced cosmic ray distribution model has been adopted and the fits are in ROI I. The left panel is extracted from the top panel of Fig.~\ref{fig:ts}.
The right panel presents the results when all the sources in 2FGL were ignored.
}
\label{fig:nops}
\end{figure}

Further more, we fit the data by ignoring all 1873 point/extended sources in 2FGL.
For illustration we have taken the Lorimer's pulsar-traced cosmic ray distribution model and the fits have been performed in ROI I. As shown in  Fig.~\ref{fig:nops}, the TS values of the additional dark matter-like GeV excess component have not been considerably changed by such a simplification. 
In view of these facts, we conclude that the point/extended sources in the region(s) of interest do not play an important role in modifying our analysis results.

\label{app:src}

\section{Likelihood method}
Here we are going to give a brief introduction to the principle of probability involved in this work.
The total number of all sky gamma-ray photons is a Poisson variable and the sequence of the number of photons in each spatial bin obeys polynominal distribution. So the numbers of photons in each spatial bin are independent Poisson variables, which can be proved as following.

If $n_t=\sum_{i=1}^N n_i$ is a Poisson variable (with expected value $\mu_t$), and $\vec{n}=(n_1,n_2,...,n_l)~(\sum_{i=1}^l n_i=n_t)$ obeys polynomial distribution, the combined distribution is equal to the product of probabilities of polynomial distribution and Poisson distribution
\begin{equation}
\begin{aligned}
P(n_1,n_2,\cdots,r_l,n_t)&=M(\vec{n};n_t,\vec{p})\cdot P(n_t;\mu_t)\\
&=\frac{n_t!}{n_1!n_2!\cdots n_l!}p_1^{n_1}p_2^{n_2}\cdots p_l^{n_l}\cdot \frac{1}{n_t!}\mu_t^{n_t}e^{-\mu_t},
\end{aligned}
\end{equation}
where $\sum_{i=1}^l p_i=1,~\sum_{i=1}^l n_i=n_t$.

So it can be refined to
\begin{equation}
P(n_1,n_2,\cdots,r_l,n_t)=P(r_1;\mu_t p_1)P(r_2;\mu_t p_2)\cdots P(r_l;\mu_t p_l),
\end{equation}
indicating that $n_1,n_2,\cdots,n_l$ are independent Poisson variables.

The Logarithm of the combined likelihood function of independent Poisson variables is defined as
\begin{equation}
\label{eq:likelihood_fuc}
\ln\mathcal{L}=\sum_i n_i\ln\mu_i-\mu_i-\ln n_i!,
\end{equation}
where $n_i$ is the observed photon counts in each spatial bin (DATA), and $\mu_i$ is the expected counts in each bin (MODEL). It is unnecessary to evaluate the term of $(-\ln n_i!)$ because it is independent of the model parameters. The error bars of model parameters are simply the square root of the diagonals of the covariance matrix.

We employ spatially binned data from Fermi-LAT as DATA. And the MODEL consists of these components: diffuse Galactic emission (DGE), isotropic background, Fermi bubbles, Dark Matter annihilation and the LAT 2-year catalog (2FGL) sources. We consider the 2FGL sources as a part of MODEL but not a simple subtraction from DATA, since the "raw data - 2FGL sources" is not a Poisson variable. The likelihood function Eq.(\ref{eq:likelihood_fuc}) is reasonable only if DATA is a Poisson variable and MODEL is the expected value of Poisson distribution.

The {\it test statistic} of the DM component is defined as
\begin{equation}
\mathrm{TS}=-2\ln\left(\frac{\mathcal{L}_\mathrm{null}}{\mathcal{L}_\mathrm{best}}\right),
\end{equation}
where $\mathcal{L}_\mathrm{null}$ is the best fit likelihood without DM component, and $\mathcal{L}_\mathrm{best}$ is the best fit likelihood with DM.
\label{app:likelihhood}

\end{appendix}
\end{document}